\documentclass[12pt, draftclsnofoot, onecolumn]{IEEEtran}

\usepackage[export]{adjustbox}
\usepackage{setspace}
\usepackage{bbm}
\usepackage{amsmath,graphics,amssymb,epsfig,subfigure,color,amsthm,cite}
\usepackage{array}
\usepackage{multirow}
\usepackage{enumerate}
\usepackage[ruled]{algorithm2e}
\SetKwInOut{Input}{Input}
\SetKwInOut{Output}{Output}
\DontPrintSemicolon

\usepackage{bm}
\usepackage[cmintegrals]{newtxmath}
\usepackage{tablefootnote}

\theoremstyle{plain}
\newtheorem{thm}{Theorem}
\newtheorem{lem}{Lemma}
\newtheorem{cor}{Corollary}
\theoremstyle{definition}

\newtheorem{rem}{Remark}
\makeatletter
\renewcommand\p@subfigure{\thefigure-}
\makeatother
\newcommand{\rom}[1]{\uppercase\expandafter{\romannumeral #1 \relax}}
\begin{document}
 \title{FDD Massive MIMO Without CSI Feedback} 

\author{\IEEEauthorblockN{Deokhwan Han,~\IEEEmembership{Student Member,~IEEE,} 
 Jeonghun Park,~\IEEEmembership{Member,~IEEE,} and \IEEEauthorblockN{Namyoon~Lee},~\IEEEmembership{Senior Member,~IEEE}}\\
 \thanks{
Deokhwan Han is with the Department of Electrical Engineering, POSTECH, Pohang, Gyeongbuk 37673, South Korea (e-mail: dhhan@postech.ac.kr).
 Jeonghun Park is with the School of Electrical Engineering, Yonsei University, Seoul, South Korea
(e-mail: jhpark@yonsei.ac.kr).
 Namyoon Lee is with School of Electrical Engineering, Korea University, South Koera (e-mail: namyoon@korea.ac.kr).}
}

\maketitle
\begin{abstract}

 Transmitter channel state information (CSIT) is indispensable for the spectral efficiency gains offered by massive multiple-input multiple-output (MIMO) systems. In a frequency-division-duplexing (FDD) massive MIMO system, CSIT is typically acquired through downlink channel estimation and user feedback, but as the number of antennas increases, the overhead for CSI training and feedback per user grows, leading to a decrease in spectral efficiency. In this paper, we show that, using uplink pilots in FDD, the downlink sum spectral efficiency gain with perfect downlink CSIT is achievable when the number of antennas at a base station is infinite under some mild channel conditions. The key idea showing our result is the mean squared error-optimal downlink channel reconstruction method using uplink pilots, which exploits the geometry reciprocity of uplink and downlink channels. We also present a robust downlink precoding method harnessing the reconstructed channel with the error covariance matrix. Our system-level simulations show that our proposed precoding method can attain comparable sum spectral efficiency to zero-forcing precoding with perfect downlink CSIT, without CSI training and feedback.

\end{abstract}

\begin{IEEEkeywords}
Frequency-division-duplexing (FDD) massive multiple-input-multiple-output (MIMO), robust downlink precoding.
\end{IEEEkeywords}

\IEEEpeerreviewmaketitle
\section{Introduction}
\subsection{Motivation}

Massive multiple-input multiple-output (MIMO) is a promising technology for achieving high spectral efficiency in future cellular networks, as demonstrated by its extensive study in time-division-duplexing (TDD) massive MIMO \cite{marzetta2010noncooperative,ngo2013energy}. Channel reciprocity is a key aspect in TDD that enables the acquisition of downlink channel state information at the transmitter without additional training or feedback, resulting in substantial gains in energy and spectral efficiencies. However, in frequency-division-duplexing (FDD), the non-reciprocal nature of the downlink and uplink channels requires additional training and feedback, leading to significant spectral efficiency losses and underrating of FDD massive MIMO in practice \cite{jindal2006mimo,love2005limited,maddah2012completely,yang2012degrees,gou2012optimal,lee2012not,lee2013space,bjornson2016massive}.



 Significant progress has been made in recent years in reducing CSI feedback overhead in FDD massive MIMO using i) compressive sensing (CS) techniques \cite{gao2015structured,han2017compressed,liang2020deep}, ii) two-stage precoding methods exploiting channel spatial correlation effects \cite{adhikary2013joint,el2014spatially,alkhateeb2014channel}, and iii) DL channel extrapolation with UL pilots \cite{wang2018deep,alrabeiah2019deep,alrabeiah2020deep,yang2020deep}. 
 In particular, estimating DL channel using UL pilots is a attracting solution that can completely eliminate the tremendous signaling overhead associated with DL channel training and CSI feedback \cite{vasisht2016eliminating,khalilsarai2018fdd}. However, accurately estimating DL channels from UL pilots is extremely challenging due to the dearth of channel reciprocity. 
  
 Notwithstanding the absence of explicit channel reciprocity, some channel parameters, including angle of arrivals (AoAs) and path gains, are still useful. This is because they are frequency-invariant, so that reciprocity partially holds for these parameters in UL and DL channels \cite{vasisht2016eliminating,khalilsarai2018fdd}. 
 Consequently, it is possible to reconstruct the DL channels with a certain level of accuracy from 
 UL pilots by exploiting these frequency-invariant channel parameters. In this paper, we make progress in this direction. In particular, our key finding is that a robust DL precoding harnessing these frequency-invariant channel parameters can provide the substantial spectral efficiency gains of the FDD massive MIMO with perfect CSIT when the number of antenna at the BS is infinite.

\subsection{Related Works}

Traditionally,  in FDD multi-user MIMO, much work in the literature has focused on characterizing the required CSI feedback overhead to attain multiplexing gains \cite{jindal2006mimo}. In a vector broadcast channel, it was shown that the feedback rate to obtain full-multiplexing gains scales with both the number of antennas at the BS and the signal-to-noise ratio (SNR) at receivers in dB \cite{jindal2006mimo}. This result has been extended in the context of multi-cell MU-MIMO settings \cite{lee2010adaptive,bhagavatula2010adaptive,park2016optimal,park2018feedback}. The conclusion drawn from these prior studies was that the CSI feedback overhead to mitigate interference effects scales with the number of antennas; thereby, this tremendous CSI feedback overhead is a significant hindrance for FDD massive MIMO.

CSI compression is necessary to diminish the feedback overhead in FDD massive MIMO. Exploiting the sparse representation of the channel in the beam domain \cite{loyka2001channel,sayeed2002deconstructing,3gpp2003spatial,brady2013beamspace}, the prior work has applied compressive sensing (CS) methods \cite{gao2015structured,han2017compressed,liang2020deep,bajwa2010compressed,alkhateeb2014mimo}. Thanks to CS theory \cite{candes2008introduction}, this approach possibly reduces the size of the effective channel vector for CSI feedback. For instance, a low-complexity sparse signal recovery algorithm \cite{tropp2007signal} allows to compress the channel vector size to the product of the log of the number of BS antennas and the sparsity level. Recently, this CS-based feedback strategy has been extended with deep learning to attain the enhanced CSI compression gain \cite{liang2020deep}. Leveraging the image compression techniques developed in deep-learning, a compressed feature of a channel matrix is obtained using convolutional neural networks (CNNs). Then, the BS rebuilds the channel matrix received from the feature matrix via feedback links. This deep-learning-aided CSI matrix compression technique can diminish the amount of CSI without an explicit sparse representation of the channel. However, it requires a large number of training samples to optimize the DNNs, and the CSI reconstruction performance is limited when the data distributions for training and testing are distinct, which is the common bottleneck of the deep-learning approach for wireless systems. These CS-based and deep-learning-aided methods can compress CSI feedback amount considerably compared to the traditional limited feedback approaches using vector quantization. However, it still requires a noticeable feedback overhead for an extremely large array size envisioned in the recent eXtreme massive MIMO 6G \cite{nokia2011extreme}, or the channel sparsity is not sufficiently advocated. 

Combined with precoding strategy, a reduced DL training and CSI feedback stategy was proposed in \cite{adhikary2013joint}. Harnessing the structure of the spatial correlation of DL channels, a two-stage precoding method called joint spatial division multiplexing (JSDM) was presented, which sequentially uses the second-order channel statistics and the instantaneous CSI in the reduced dimensional space. It turned out that JSDM can achieve the optimal spectral efficiency attained with full CSIT under some mild conditions. This idea has also been extended in the context of mmWave massive MIMO, in which hybrid precoding that concatenates analog precoding with digital precoding can reduce both DL channel training and CSI feedback overhead from the two-stage precoding architecture \cite{el2014spatially,alkhateeb2014channel}.

The most relevant prior work to this paper is to use UL pilots to acquire DL CSIT in FDD massive MIMO \cite{vasisht2016eliminating, khalilsarai2018fdd}. Using the fact that the angular scattering functions between UL and DL channels are identical over frequency, i.e., frequency-invariant, a BS can accurately estimate the DL channel covariance matrix from UL pilots. The DL transmission method using the covariance matrix was proposed in \cite{khalilsarai2018fdd} and showed the spectral efficiency gain over the CS-aided CSI feedback strategy. The DL channel reconstruction algorithm was proposed using UL pilots in generic FDD wireless systems \cite{vasisht2016eliminating}. The idea was to reconstruct DL channels using frequency-invariant channel parameters, including AoAs and path gains, between UL and DL channels. This frequency extrapolation method has been further extended in \cite{rottenberg2019channel,rottenberg2020performance}, where a high-resolution channel estimation technique was presented with a theoretical lower bound of the extrapolated channel. To reduce the beam training overhead in mmWave systems, it has been shown that exploiting channel knowledge of sub-6GHz bands is useful \cite{ali2017millimeter,maschietti2019coordinated}. Recently, with the power of deep learning to identify a non-linear transform mapping, the DL channel reconstruction problem from UL pilots has been extensively tackled in \cite{wang2018deep,alrabeiah2019deep,alrabeiah2020deep,yang2020deep}. For instance, a DL channel estimation technique was presented in  \cite{alkhateeb2019deepmimo} using DeepMIMO datasets. The deep-learning-aided approach has opened a new opportunity to find the complicated non-linear mapping function for the DL channel reconstruction. This approach, however, is vulnerable to changing channel distributions between training and testing. The frequency extrapolation methods \cite{vasisht2016eliminating,rottenberg2019channel,rottenberg2020performance,ali2017millimeter,maschietti2019coordinated} are also limited to accurately estimating frequency-variant parameters such as the phase information of individual channel paths. These parameters with uncertainty hinders to accurately estimate instantaneous DL CSIT. In addition, all aforementioned works do not show how to optimally exploit the partial channel reciprocity of frequency-invariant parameters for DL precoding. 

\subsection{Contributions}


\begin{itemize}
    \item We present a DL channel reconstruction method that utilizes UL pilots. Our DL reconstruction method consists of three steps. Firstly, the BS performs UL channel estimation on each user through the use of orthogonal UL pilots assigned to the users. In the second stage, the BS estimates frequency-invariant channel parameters, including AoAs and path gains. Finally, the BS joinlty uses the estimated UL channel information, along with the AoAs and path gains, and the uniform linear array (ULA) structure at the BS to reconstruct the DL channel vector with unknown phase information for each channel path. Based on the assumption of a uniform phase distribution, the worst-case scenario, we present the MSE-optimal DL channel reconstruction methods. We highlight that our approach differs from previous works in \cite{vasisht2016eliminating,rottenberg2019channel}. Our methods utilize the UL channel knowledge along with the channel parameters jointly to minimize the DL channel reconstruction error. 

\item To gauge the exactness of the proposed DL channel reconstruction method, we derive the MSE matrix in closed form. From the derived MSE matrix, we elucidate that the reconstruction error is a function of the ratio between UL and DL frequencies; a larger ratio makes more reconstruction error, which confirms our intuition. More importantly, from the MSE analysis, we prove that the outer product between the actual instantaneous DL channel can be tightly approximated as the sum of the outer product of the reconstructed DL channels and the MSE matrix in the infinite number of antennas. 

\item We also propose a robust DL precoding algorithm that takes into account both the reconstructed DL channels and the reconstruction error covariance matrix. This robust precoding method is an extension of our previous work in \cite{choi2019joint}, which guarantees to find a local optiaml solution for a joint user selection, power control, and beamforming to maximize the downlink sum spectral efficiency with incomplete CSIT. The key difference in our current work is that we have derived the second order local optimality condition for the sum spectral efficiency maximization problem to ensure the local optimality of the general power iteration precoding (GPIP) algorithm.

\item System level simulation results reveal that the proposed DL precoding algorithm with the DL channel reconstruction can achieve the comparable ergodic DL spectral efficiency to the zero-forcing (ZF) precoding with perfect CSIT, even with not-so-many BS antennas. This result confirms that the the FDD massive MIMO gains can be achieved using UL pilots, provided that the AoAs and path gains of the channel are available.
    

    
\end{itemize}

 \subsection{Paper Organization and Notations}

 This paper is organized as follows: Section \rom{2} covers the network topology and channel model. Section \rom{3} outlines the proposed DL channel reconstruction framework with partial geometric parameters. Section \rom{4} details the robust DL data transmission framework. Section \rom{5} establishes local optimality conditions for the formulated optimization problems. Section \rom{6} presents an efficient algorithm for finding local optimal solutions. Numerical results are in Section \rom{7}. Conclusion in Section \rom{8}.

In this paper, we use the following notations: $\mathbb{C}$ represents the set of complex numbers, $\mathbb{R}$ represents the set of real numbers, $\otimes$ denotes the Kronecker product, $\odot$ denotes the Hadamard product, and ${\sf Tr}\left\{\cdot\right\}$ is the matrix trace operation. The expectation of a random vector ${\bf x}$ is represented by $\mathbb{E}\left[{\bf x}\right]$. The real part of a complex scalar $x$ is denoted by ${\rm Re}\left[ x \right]$. The minimum and maximum eigenvalues of a matrix ${\bf A}$ are represented by $\rho_{\sf min}\left({\bf A}\right)$ and $\rho_{\sf max}\left({\bf A}\right)$, respectively. An $N\times N$ identity matrix is represented by ${\bf I}_N$. A complex Gaussian distribution with mean vector ${\bf m}$ and covariance matrix ${\bf R}$ is represented by ${\bf x}\sim\mathcal{CN\left({\bf m},{\bf R}\right)}$. The column vectors of matrix ${\bf A}\in\mathbb{C}^{N\times N}$ are represented by $\left\{ {\bf a}_1,\cdots,{\bf a}_N\right\}$, and column vector ${\bf a}\in\mathbb{C}^{N\times 1}$ consists of elements $\left\{{a}_1,\cdots,{a}_N\right\}$.

\section{System Model}
In this section, we explain an FDD massive MIMO system model where the BS has $N$ antennas and there are $K$ users each with a single receive antenna. For ease of exposition, we shall focus on a single-cell network to clearly explain the proposed methods. Nonetheless, our method can be readily applicable to multi-cell networks by incorporating pilot contamination effects with a minor modification. We will verify the effectiveness of the proposed methods in both single-cell and multi-cell networks in Section \ref{sec:Simulation_Results}.  



\subsection{UL and DL Channel Models}
\begin{figure}[t]
	\centering
    \includegraphics[width=0.5\linewidth]{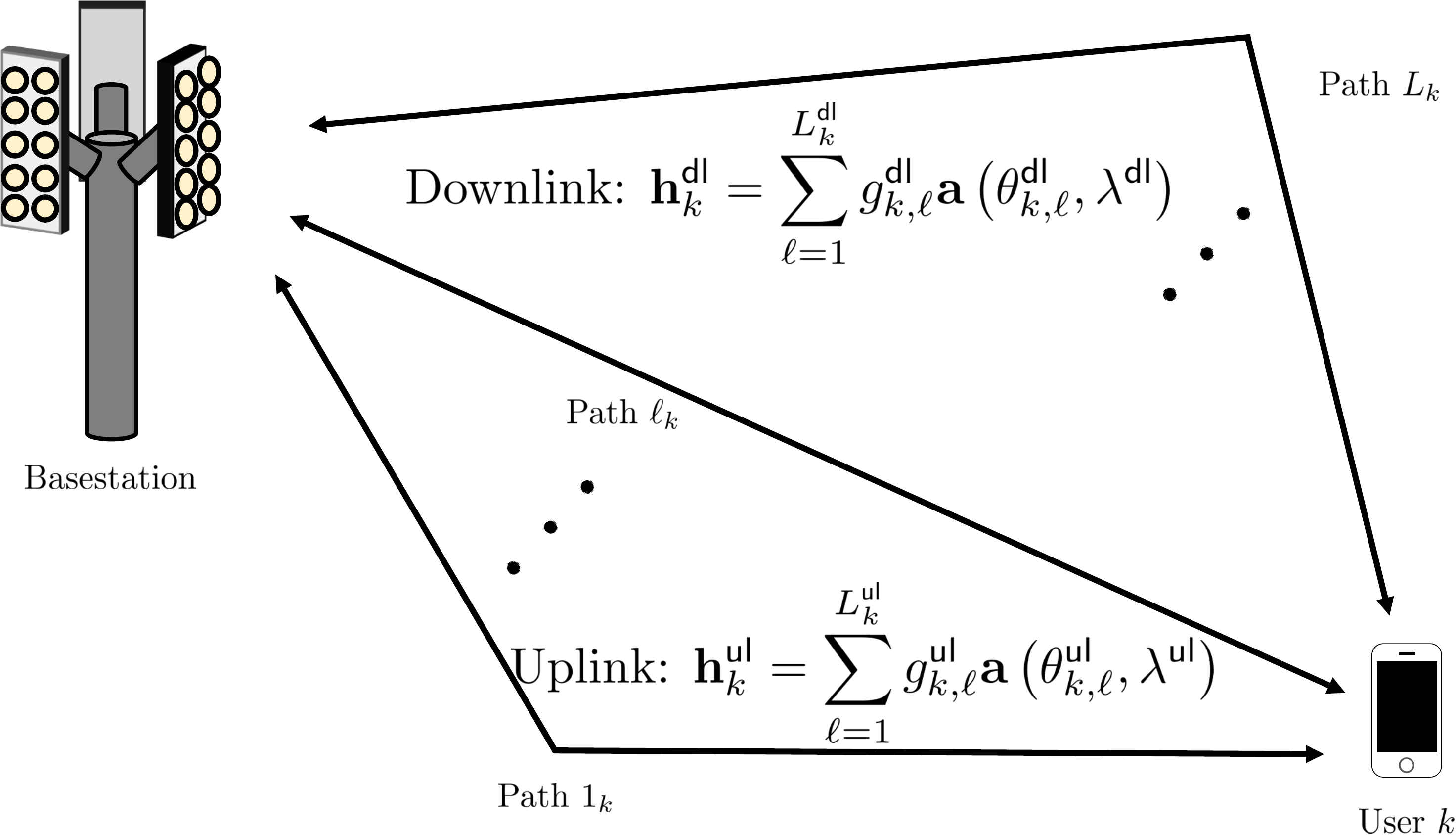}
  \caption{The UL and DL multi-path channel models with ULA at the BS.} \label{fig_System_Model}
\end{figure}

{\bf UL channel model:}  We commence with defining the UL channel from user $k\in [K]$ to the BS. Let ${\bf h}_k^{{\sf ul}}\in \mathbb{C}^N$ be the UL channel vector defined as the sum of array response vectors. Assuming uniform-linear-array (ULA) antenna elements at the BS, the narrowband UL channel model is
\begin{align}
    {\bf h}_k^{{\sf ul}} = \sum_{\ell=1}^{L_k^{\sf ul}}g_{k,\ell}^{\sf ul}{\bf a}\left(\theta_{k,\ell}^{\sf ul}, \lambda^{\sf ul}\right),
    \label{eq:UL_Channel}
\end{align}
where $L_{k}^{\sf ul}\in \mathbb{Z}^{+}$ and $g_{k,\ell}^{\sf ul}\in\mathbb{C}$ denote the number of multi-paths and the complex coefficient for fading. Under the far-field assumption, the array response vector is defined as a function of angle of arrivals (AoAs) $\theta_{k,\ell}^{\sf ul}$ as
\begin{align}
    {\bf a}(\theta_{k,\ell}^{\sf ul}, \lambda^{\sf ul})=\left[1,e^{-j\frac{2\pi}{\lambda^{\sf ul}}d\sin{\theta_{k,\ell}^{\sf ul}}},\hdots,e^{-j\frac{2\pi}{\lambda^{\sf ul}}(N-1)d\sin{\theta_{k,\ell}^{\sf ul}}}\right]^{\sf T},    
\end{align}
where $\lambda^{\sf ul}$ is the wavelength and $d$ is the inter-antenna spacing. The narrowband channel of signal traversing the $\ell$th path is given by \cite{tse2005fundamentals}: 
\begin{align}
    g_{k,\ell}^{\sf ul} = b_{\ell,k}^{\sf ul}e^{-j\frac{2\pi}{\lambda^{\sf ul}}r_{k,\ell}+\phi_{k,\ell}^{\sf ul}},
\end{align}
where $b_{\ell,k}^{\sf ul}\in \mathbb{R}^{+}$ is the $\ell$th path attenuation of the UL channel from user $k$ to the BS, $r_{\ell,k}$ is the path distance, and $\phi_{\ell,k}^{\sf ul}$ is a random phase which is independent of wavelength, and it is uniformly distributed over $\left[0,2\pi\right)$ to capture small-scale fading effects by path reflections.


{\bf DL channel model:} Analogy to the UL channel model in \eqref{eq:UL_Channel}, the DL channel from the BS to user $k$ is modeled by the sum of the complex channel gains $g_{k,\ell}^{\sf dl}$ and array response vectors ${\bf a}(\theta_{k,\ell}^{\sf dl},\lambda^{\sf dl})$ representing the array phase profile corresponding to  angle of departure (AoD) $\theta_{k,\ell}^{\sf dl}$ as
\begin{align}
    {\bf h}_k^{{\sf dl}} = \sum_{\ell=1}^{L_k^{\sf dl}}g_{k,\ell}^{\sf dl}{\bf a}(\theta_{k,\ell}^{\sf dl}, \lambda^{\sf dl}),
    \label{eq:DL_Channel}
\end{align}
where $L_k^{\sf dl}$ is the number of multi-paths for the DL channel and the array response vector ${\bf a}(\theta_{k,\ell}^{\sf dl},\lambda^{\sf dl})$ is defined as
\begin{align}
    {\bf a}(\theta_{k,\ell}^{\sf dl}, \lambda^{\sf dl})=\left[1,e^{-j\frac{2\pi}{\lambda^{\sf dl}}d\sin{\theta_{k,\ell}^{\sf dl}}},\hdots,e^{-j\frac{2\pi}{\lambda^{\sf dl}}(N-1)d\sin{\theta_{k,\ell}^{\sf dl}}}\right],   
\end{align}
where $\lambda^{\sf dl}$ is the DL wavelength. The complex channel coefficient of the $\ell$th path is similarly defined as
\begin{align}
    g_{k,\ell}^{\sf dl} = b_{\ell,k}^{\sf dl}e^{-j\frac{2\pi}{\lambda^{\sf dl}}r_{k,\ell}+\phi_{k,\ell}^{\sf dl}}.
\end{align}
where $b_{\ell,k}^{\sf dl}\in \mathbb{R}^{+}$ is the $\ell$th path attenuation of the DL channel from the BS to user $k$.  We also assume that random phase of the DL channel path $\phi_{k,\ell}^{\sf dl}$ is uniformly distributed over $[0,2\pi)$.

Our simplified single-cluster channel models, represented in \eqref{eq:UL_Channel} and \eqref{eq:DL_Channel}, capture the scattering effects from diffuse reflections within a cluster through multiple paths. These multi-path channel models have widely used in the literature {\cite{3gpp2003spatial}}. Our model assumes that the cluster has multiple dominant paths displaying macro-level channel propagation effect, with angle spreads concentrated at the AoA and AoD in each path. This channel model has empirically proven accurate especially in high frequencies, where scattering from diffuse reflections is less pronounced.

\subsection{Frequency-invariant channel parameters}
It is clear that the UL and DL channels, which operate at different wavelengths, $\lambda^{\sf ul}$ and $\lambda^{\sf dl}$, are not reciprocal. This is due to the fact that the channel coefficients, $ g_{k,\ell}^{\sf ul} $ and $ g_{k,\ell}^{\sf dl} $, and array response vectors, ${\bf a}(\theta_{k,\ell}^{\sf ul},\lambda^{\sf ul})$ and ${\bf a}(\theta_{k,\ell}^{\sf dl},\lambda^{\sf dl})$, vary with the wavelength. However, a significant observation is that some critical geometric parameters in the UL and DL channels are commonly frequency-invariant. These frequency-invariant parameters include:
\begin{itemize}
    \item The angles of arrival (AoAs), denoted by $\theta_{k,\ell}^{\sf ul}$, and angles of departure (AoDs), represented by $\theta_{k,\ell}^{\sf dl}$, are fundamental parameters in the wave propagation. These parameters are often assumed to be identical, due to the inherent symmetry of the antenna and path geometry.
 
    \item The channel attenuation parameters, $ b_{k,\ell}^{\sf ul}$ and $ b_{k,\ell}^{\sf dl}$, are frequency-dependent as modeled in \cite{maccartney2017rural,samimi2015probabilistic}. This is due to the fact that the path loss with omni-directional antennas is modeled as follows:
\begin{align}
-10\log{b_{k,\ell}^{\sf ul}}=20\log{\left(\frac{4\pi r_0}{\lambda^{\sf ul}}\right)}+m\log{\left(\frac{r_{k,\ell}}{r_0}\right)}+X_{k,\ell}.
\end{align}
The parameters $r_0$, $m$, and $X_{k,\ell}$ denote the reference distance, path-loss exponent, and shadowing parameter, respectively. As the path distance $r_{k,\ell}$ is shared between the UL and DL channels, the dominant term, $m\log{\left(\frac{r_{k,\ell}}{r_0}\right)}$, becomes identical in both channels. Hence, the path attenuation terms in the UL and DL channels can be considered frequency-invariant, as stated in \cite{vasisht2016eliminating}. Thus, we can assume $ b_{k,\ell}^{\sf ul}\simeq b_{k,\ell}^{\sf dl}$.
    \item The phase variations, $\phi_{k,\ell}^{\sf ul}$ and $\phi_{k,\ell}^{\sf dl}$, arise from phenomena such as reflection and refraction. As a result, these parameters are consistent between the UL and DL so long as the physical environment remains unchanged.
\end{itemize}

In this paper, we utilize the frequency-invariant channel parameters to recreate the DL channels through uplink pilot signals. For ease of notation, we simplify the AoAs and DoAs by removing the superscript, $\theta_{k,\ell} = \theta_{k,\ell}^{\sf ul} = \theta_{k,\ell}^{\sf dl}$. The same holds for the path attenuation $b_{k,\ell}=b_{k,\ell}^{\sf ul}=b_{k,\ell}^{\sf dl}$ and the phase variation $\phi_{k,\ell}=\phi_{k,\ell}^{\sf ul}=\phi_{k,\ell}^{\sf dl}$. It is also important to note that we assume that the number of channel paths in the UL and DL are equal, i.e., $L_k=L_{k}^{\sf ul}= L_{k}^{\sf dl}$.

\section{MSE-Optimal DL Channel Reconstruction}
In this section, we present the MSE-optimal channel reconstruction algorithms for the DL channel, assuming perfect knowledge of both the UL channels ${\bf h}_{k}^{\sf ul}$ and partial frequency-invariant channel parameters $\left\{\theta_{k,\ell}, b_{k,\ell}\right\}_{\ell=1}^{L_k}$ for each user $k\in[K]$ for ease of exposition. We will show the effect of imperfect knowledge of UL channel and the parameters on the DL performance in Section V-A. We further analyze the DL channel reconstruction error covariance matrix to quantify the accuracy of the DL channel reconstruction as a function of the UL and DL carrier frequencies.

\subsection{MSE-Optimal DL Channel Reconstruction}

We commence by introducing a useful lemma that to derive the MSE-optimal DL channel reconstruction. 

\begin{lem} \label{lem1}
Suppose the phases of UL and DL channel paths are uniformly distributed, i.e., $\angle{g_{k,\ell}^{\sf ul}} =\frac{2\pi}{\lambda^{\sf ul}}r_{k,\ell}-\phi_{k,\ell} \in\left[0,2\pi\right)$ for $\ell\in L_k$ and $\angle{g_{k,\ell}^{\sf dl}} =\frac{2\pi}{\lambda^{\sf dl}}r_{k,\ell}-\phi_{k,\ell} \in\left[0,2\pi\right)$ for $\ell\in L_k$. Then, the correlation between ${\bf g}_k^{\sf ul}$ and ${\bf g}_k^{\sf dl}$ conditioned on $ b_{k,1},\ldots, b_{k,L_k}$ is given by
\begin{align}
    \mathbb{E}\left[ {\bf g}_k^{\sf dl}\left({\bf g}_k^{\sf ul}\right)^{\sf H} | b_{k,1},\ldots, b_{k,L_k}\right] = \eta{\bf \Sigma}_k,\quad\forall k\in[K], \label{eq:Corr_UL_DL}
\end{align}
where 
\begin{align}
    {\bf \Sigma}_k &= {\sf diag}\left(\left[b_{k,1}^2,\cdots,b_{k,L_k}^2\right]\right) ,\nonumber\\
    \eta&=\frac{1}{2\pi(\frac{\lambda^{\sf ul}}{\lambda^{\sf dl}}-1)}\left(\sin\left(2\pi\frac{\lambda^{\sf ul}}{\lambda^{\sf dl}}\right)-2j\sin^2\left(\pi\frac{\lambda^{\sf ul}}{\lambda^{\sf dl}}\right)\right).\label{eq:eta}
\end{align} 
\end{lem} 
\begin{IEEEproof}
    See the Appendix \ref{Proof_for_Corr_UL_DL}.
\end{IEEEproof}

\vspace{0.1cm}
Lemma \ref{lem1} implies that the correlation matrix $\eta{\bf \Sigma}_k$ is a scaled version of ${\bf \Sigma}_k$ by a factor of $\eta$. To understand the effect of $\eta$, we need to analyze its behavior with the changes in the carrier frequency gap between UL and DL. For instance, if the carrier frequency in UL and DL are equal, i.e., $f_c^{\sf ul} = f_c^{\sf dl}$, the correlation simplifies to the path attenuation power ${\bf \Sigma}_k$ since $\eta=1$.  On the other hand, if $f_c^{\sf ul} \ne f_c^{\sf dl}$, $\eta$ reflects the deviation as the ratio of $\frac{\lambda^{\sf ul}}{\lambda^{\sf dl}}$.


Using this property, we show how to recreate the DL channels through the utilization of the UL channels and partial geometric parameters. We first assume that the BS has a perfect knowledge of the UL channels and partial geometric parameters, represented as $\left\{{\bf h}_k^{\sf ul},{\bf\Theta}_k,{\bf \Sigma}_k\right\}$ for ${k\in[K]}$.

\begin{thm}\label{thm:OptimalDlChannelEstimator} 
The MSE-optimal DL channel estimate is given by
\begin{align}
    {\bf\hat h}_k^{\sf dl, MMSE} 
    &= \eta{\bf A}_k^{\sf dl}{\bf \Sigma}_k^{\frac{1}{2}}\left({\bf \Sigma}_k^{-\frac{1}{2}}\left({\bf A}_k^{\sf ul}\right)^{\dagger}{\bf h}_k^{\sf ul}\right)^{\frac{\lambda^{\sf ul}}{\lambda^{\sf dl}}},\label{eq: Optimal_Estimator}
\end{align}
where ${\bf A}_k^{\sf ul}=\left[{\bf a}(\theta_{k,1},\lambda^{\sf ul}), \ldots , {\bf a}(\theta_{k,L_k},\lambda^{\sf ul})\right]\in \mathbb{C}^{N\times L_k}$ and ${\bf A}_k^{\sf dl}=\left[{\bf a}(\theta_{k,1},\lambda^{\sf dl}), \ldots , {\bf a}(\theta_{k,L_k},\lambda^{\sf dl})\right]\in \mathbb{C}^{N\times L_k}$ are the UL and DL array response matrices, respectively.
 
\end{thm}
\begin{IEEEproof}
    See Appendix \ref{Proof_for_OptimalDlChannelEstimator}.
\end{IEEEproof}

\vspace{0.1cm}
Theorem \ref{thm:OptimalDlChannelEstimator} demonstrates that the MSE-optimal DL channel estimate is derived from a combination of four elements: the UL channel ${\bf h}_k^{\sf ul}$, the UL array response matrix ${\bf A}_k^{\sf ul}$, the correlation matrix from Lemma 1 ($\eta {\bf \Sigma}_k$), and the DL array response matrix ${\bf A}_k^{\sf dl}$. The estimation process starts with the estimation of the UL channel ${\bf h}_k^{\sf ul}$. Then, it undergoes a transformation using $\frac{\lambda^{\sf ul}}{\lambda^{\sf dl}}$ as the exponent and ${\bf \Sigma}_k^{-\frac{1}{2}}\left({\bf A}_k^{\sf ul}\right)^{\dagger}$. Finally, the DL channel estimate is obtained by projecting the result onto the DL array matrix through multiplication with the normalization matrix $\eta{\bf A}_k^{\sf dl}{\bf \Sigma}_k^{\frac{1}{2}}$.


Next, we present a linear-MMSE (L-MMSE) DL channel estimation in the following corollary.

\begin{cor}\label{thm:LinearDlChannelEstimator}
The MSE-optimal DL channel estimate under the linear map constraint is given by
\begin{align}
 {\hat{\bf h}}_{k}^{\sf dl, L-MMSE}  
   ={\rm Re}\{\eta\}{\bf A}_k^{\sf dl}\left({\bf A}_k^{\sf ul}\right)^{\dagger} {\bf h}_k^{\sf ul}.\label{eq:col1}
\end{align}
where ${\rm Re}\{\eta\}= \frac{1}{2\pi(\frac{\lambda^{\sf ul}}{\lambda^{\sf dl}}-1)} \sin\left(2\pi\frac{\lambda^{\sf ul}}{\lambda^{\sf dl}}\right). $
\end{cor}
\begin{IEEEproof}
See Appendix \ref{Proof_for_Corr1}.
\end{IEEEproof}

The L-MMSE DL channel estimation method differs from the MSE-optimal DL channel estimate in \eqref{eq: Optimal_Estimator} as it requires information about the UL array response matrix ${\bf A}_k^{\sf ul}$, the DL array response matrix ${\bf A}_k^{\sf dl}$, the real part of the carrier normalization constant ${\rm Re}\{{\eta}\}$, and the UL channel ${\bf h}_k^{\sf ul}$. However, the channel path gain information ${\bf \Sigma}_k$ is not necessary for the L-MMSE estimation, simplifying the DL channel reconstruction process. Furthermore, the L-MMSE estimation uses the real part of $\eta$ in \eqref{eq:eta} as its carrier normalization term.


\subsection{MSE Analysis}
To evaluate the precision of the DL channel reconstruction, we assess the MSE performance of the proposed DL channel reconstruction method in \eqref{eq: Optimal_Estimator} and \eqref{eq:col1}. We first construct the MSE matrix of each estimator and derive the MSE by using MSE matrix. 
\begin{thm}\label{thm:MseMatrxOfOptimal}
Let ${\bf e}_k={\bf h}_k^{\sf dl}-{\bf\hat h}_k^{\sf dl, MMSE}$ be the DL channel reconstruction error. Then, the error covariance matrix with knowledge of ${\bf h}_k^{\sf ul},{\bf\Theta}_k$, and ${\bf \Sigma}_k$ is given by is 
\begin{align}
    {\bf \Phi}_k^{\sf MMSE}&=\mathbb{E}\left[\left.{\bf e}_k{\bf e}_k^{\sf H}\right|{\bf h}_k^{\sf ul},{\bf\Theta}_k,{\bf \Sigma}_k\right]\nonumber\\
    &=\left(1-|\eta|^2\right){\bf A}_k^{\sf dl}{\bf \Sigma}_k\left({\bf A}_k^{\sf dl}\right)^{\sf H}.    
\end{align}
\end{thm}
\begin{IEEEproof}
    See Appendix \ref{ProofForThm2}.
\end{IEEEproof}

The following corollary states the MSE matrix when using the L-MMSE DL channel reconstruction method in \eqref{eq:col1}. 
\begin{cor}\label{thm:MseMatrixOfLMSE}
Let ${\bf \tilde e}_k={\bf h}_k^{\sf dl}-{\bf\hat h}_k^{\sf dl, L-MMSE}$ be the DL channel reconstruction error. Then, the error covariance matrix with knowledge of ${\bf h}_k^{\sf ul},{\bf\Theta}_k$, and ${\bf \Sigma}_k$ is given by is 
\begin{align}
    {\bf \Phi}_k^{\sf L-MMSE}
    &=\mathbb{E}\left[\left. {\bf \tilde e}_k{\bf \tilde e}_k^{\sf H}\right|{\bf h}_k^{\sf ul},{\bf\Theta}_k,{\bf \Sigma}_k\right]\nonumber\\
    &=\left(1-{\rm Re}\{\eta\}^2\right){\bf A}_k^{\sf dl}{\bf \Sigma}_k\left({\bf A}_k^{\sf dl}\right)^{\sf H}.    
\end{align}
\end{cor}
\begin{IEEEproof}
    The proof is straight from Appendix \ref{ProofForThm2}.
\end{IEEEproof}

Theorems \ref{thm:MseMatrxOfOptimal} and Corollary \ref{thm:MseMatrixOfLMSE} demonstrate that the MSE matrix is influenced by two elements: i) the DL array response matrix ${\bf A}_k^{\sf dl}$ and ii) the correlation matrix of the UL and DL path gain, $\eta{\bf \Sigma}_k$, as stated in Lemma \ref{lem1}. It is important to note that the MSE matrix changes based on the UL and DL frequency ratio $\frac{\lambda^{\sf ul}}{\lambda^{\sf dl}}$ as $\eta$ in \eqref{eq:eta}. To further comprehend this impact, we calculate the asymptotic MSE values for the proposed DL reconstruction algorithms, as stated in the following corollary.

\begin{figure}[t]
\subfigure[]{
\includegraphics[width=0.49\linewidth]{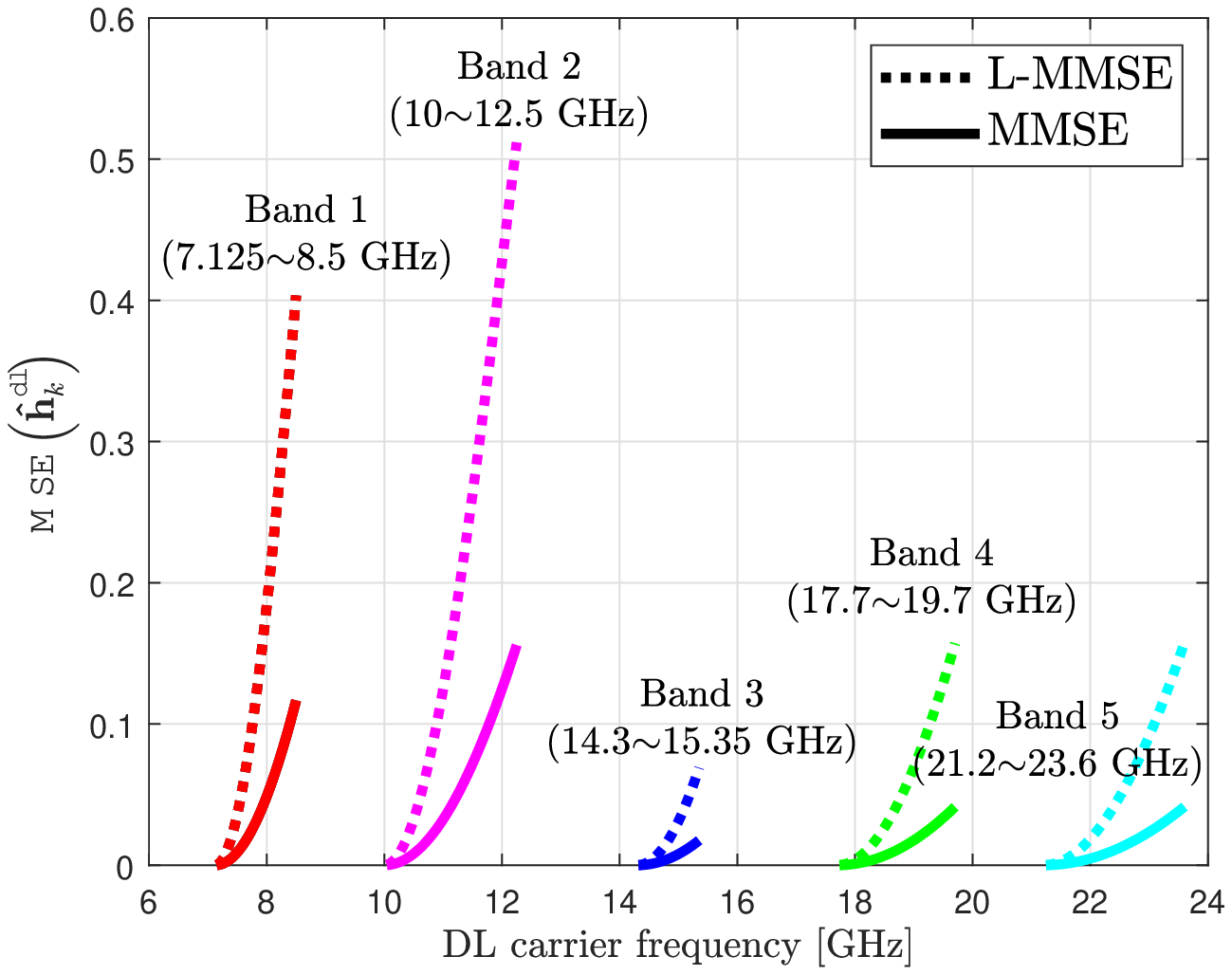}%
\label{fig_Normalized_MSE_upper_mid_bands_Optimal}
}
\subfigure[]{
\includegraphics[width=0.49\linewidth]{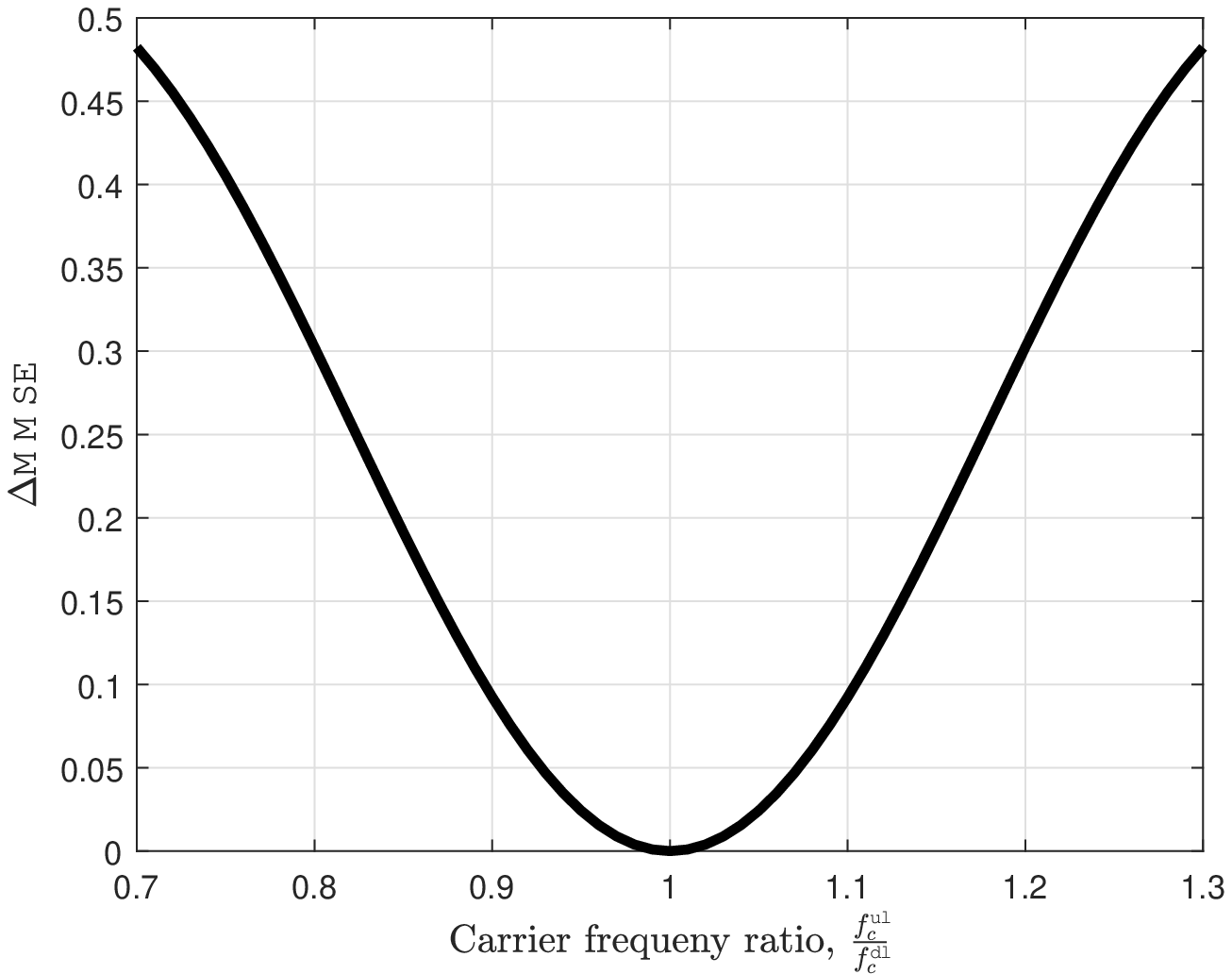}%
\label{fig_Normalized_MSE_Gap_with_Optimal}
}
\caption{(a) The MSE with increasing DL carrier frequency over upper-mid bands and (b) the normalized MSE gap between the nonlinear MMSE and the L-MMSE estimators in terms of the carrier frequency ratio $\frac{f_c^{\sf ul}}{f_c^{\sf dl}}$.
}
\label{fig_Normalized_MSE}
\end{figure}

\begin{cor}\label{cor:MMSE_Derivation} When using the MSE-optimal DL channel estimator in \eqref{eq: Optimal_Estimator}, the corresponding MSE is given by
    \begin{align}
        {\sf MSE}
        &=\lim_{N\rightarrow \infty}\frac{1}{N}{\rm Tr}\left[ {\bf \Phi}_k^{\sf MMSE}\right] /{\rm Tr}\left[ {\bf \Sigma}_k\right]\nonumber\\
        &=1-|\eta|^2.\label{eq:MSE_OptimalEstimator}
    \end{align}
    In addition, when using the linear MSE-optimal DL channel estimator in \eqref{eq:col1}, the corresponding MSE is given by
    \begin{align}
                {\sf L\text{-}MSE}
                &=\lim_{N\rightarrow \infty}\frac{1}{N}{\rm Tr}\left[ {\bf \Phi}_k^{\sf L-MMSE}\right] /{\rm Tr}\left[ {\bf \Sigma}_k\right]\nonumber\\
        &=1-{\rm Re}\{\eta\}^2.\label{eq:MSE_LinearEstimator}
    \end{align}
\end{cor}
\begin{IEEEproof}
    See Appendix \ref{Proof_for_Cor2}.
\end{IEEEproof}
Corollary \ref{cor:MMSE_Derivation} shows that both MSE and L-MSE deteriorate as the frequency gap increases. Fig. \ref{fig_Normalized_MSE_upper_mid_bands_Optimal} shows both the MSE (solid lines) and L-MSE (dotted lines) for different carrier frequencies in the upper-mid bands (7-23 GHz). In the simulation, the UL carrier frequency was fixed at 7.125, 10, 14.3, 17.7, or 21.2 GHz, and the DL carrier frequency was varied. Our results demonsrate that both MSE and L-MSE increase with the frequency gap, indicating that DL channel reconstruction from UL channel information is limited as the frequency gap grows.

We also compute the difference between L-MSE and MSE in terms of the
carrier frequency ratio $\kappa$ as
\begin{align}
    {\Delta{\sf MSE}}&=\left|\eta\right|^2-{\rm Re}\{\eta\}^2\nonumber\\&=\frac{\sin^4\left(\pi\kappa\right)}{\pi^2\left(1-\kappa\right)^2}.
\end{align}
The results shown in Fig. \ref{fig_Normalized_MSE_Gap_with_Optimal} indicate that within a 10$\%$ frequency deviation of $0.9\le\frac{f_c^{\sf ul}}{f_c^{\sf dl}}\le1.1$, the linear MSE-optimal DL channel estimator only causes a slight decrease in MSE compared to the optimal estimator. This deviation, which is commonly observed in practice, means that the use of the linear MSE-optimal DL channel estimator allows for a simplified implementation without sacrificing performance. We shall focus on the L-MMSE DL channel estimator in  \eqref{eq:col1} in the sequel. For the national simplicity, we will replace ${\bf \hat h}_k^{\sf dl, L-MMSE}$ ${\bf\Phi}_k^{\sf L-MMSE}$ by ${\bf \hat h}_k^{\sf dl}$ and ${\bf \Phi}_k$, respectively hereafter.

 \section{Robust DL Precoding}
In this section, we present a robust precoding algorithm that maximizes the sum-spectral efficiency using the DL channel reconstructed from the UL channel, ${\bf \hat h}_k^{\sf dl}$, and its MSE matrix, ${\bf \Phi}_k$. The algorithm begins by reviewing the sum-spectral efficiency maximization problem with imperfect CSIT (as introduced in \cite{choi2019joint}) and then presents a modified version taking into account ${\bf \hat h}_k^{\sf dl}$ and ${\bf \Phi}_k$ for robust DL precoding.

\subsection{Sum Spectral Efficiency Maximization}
Let us denote the downlink data symbol for user $k$ as $s_k\in\mathbb{C}$, and the corresponding precoding vector as ${\bf f}_k\in\mathbb{C}^{N\times 1}$. Under the assumption of Gaussian signaling, where $s_k$ follows a complex Gaussian distribution with mean 0 and variance $P$, the downlink transmit signal ${\bf x}\in\mathbb{C}^{N\times 1}$ can be expressed as a linear combination of $s_k$ and ${\bf f}_k$. That is,
\begin{align}
    {\bf x}=\sum_{k=1}^{K}{\bf f}_ks_k.
\end{align}
The received signal at user $k$ is,
\begin{align}
    y_k = \left({\bf h}_k^{\sf dl}\right)^{\sf H}{\bf f}_ks_k
    +\left({\bf h}_k^{\sf dl}\right)^{\sf H}\sum_{i\ne k}{\bf f}_is_i
    +z_k,
\end{align}
where $z_k\sim\mathcal{CN}(0,\sigma_k^2)$ is the complex Gaussian noise. Then, the signal-to-interference-plus-noise ratio (SINR) of user $k$ is given by 
\begin{align}
    {\sf SINR}_k 
    &= \frac{\left|\left({\bf h}_k^{\sf dl}\right)^{\sf H}{\bf f}_k\right|^2}{\sum_{i\ne k}\left|\left({\bf h}_k^{\sf dl}\right)^{\sf H}{\bf f}_i\right|^2+\sigma_k^2/P}=\frac{\sum_{i=1}^{K}{\bf f}_{i}^{\sf H}{\bf h}_{k}^{\sf dl}\left({\bf h}_k^{\sf dl}\right)^{\sf H}{\bf f}_{i}+{\sigma}_k^2/P}{\sum_{i\ne k}^{K}{\bf f}_{i}^{\sf H}{\bf h}_{k}^{\sf dl}\left({\bf h}_k^{\sf dl}\right)^{\sf H}{\bf f}_{i}+{\sigma}_k^2/P}.
\end{align}
With perfect CSIT, the DL sum-spectral efficiency is defined as
\begin{align}
    R({\bf f}_1,\ldots, {\bf f}_K) &=   \sum_{k=1}^K\log_2\left(1+{\sf SINR}_k\right) \nonumber\\
    &=\log_2\left( \prod_{k=1}^{K}\frac{\sum_{i=1}^{K}{\bf f}_{i}^{\sf H}{\bf h}_{k}^{\sf dl}\left({\bf h}_k^{\sf dl}\right)^{\sf H}{\bf f}_{i}+{\sigma}_k^2/P}{\sum_{i\ne k}^{K}{\bf f}_{i}^{\sf H}{\bf h}_{k}^{\sf dl}\left({\bf h}_k^{\sf dl}\right)^{\sf H}{\bf f}_{i}+{\sigma}_k^2/P}\right). \label{eq:sumrate}
\end{align}
Under the sum-power constraint $\sum_{k=1}^{K}\left\|{\bf f}_k\right\|_2^2 = 1$, the sum-spectral efficiency maximization problem is given by
\begin{align}
    \mathscr{P}^1:~&{\underset{\left\{{\bf f}_k\right\}_{k=1}^K\in \mathbb{C}^{N\times 1}}{\text{arg~max}}}~
     R({\bf f}_1,\ldots, {\bf f}_K)\nonumber\\
    &~\text{subject to} \sum_{k=1}^{K}\left\|{\bf f}_k\right\|_2^2 = 1.\label{eq:opt1}
\end{align}

\subsection{ Imperfect CSIT for Robust Precoding}
To solve the sum-spectral maximization problem in \eqref{eq:opt1}, either the DL channel ${\bf h}_{k}^{\sf dl}$ or its outer-product ${\bf h}_{k}^{\sf dl}\left({\bf h}_{k}^{\sf dl}\right)^{\sf H}$ must be known for each $k\in[K]$. However, in FDD, the exact outer-product ${\bf h}_{k}^{\sf dl}\left({\bf h}_{k}^{\sf dl}\right)^{\sf H}$ is unavailable and an approximation must be used for DL precoding. The following theorem shows the asymptotic error between ${\bf h}_{k}^{\sf dl}\left({\bf h}_{k}^{\sf dl}\right)^{\sf H}$ and the reconstructed CSIT using ${\bf \hat h}_{k}^{\sf dl}$ and ${\bf \Phi}_{k}$.

\begin{thm}\label{thm3}
Using the reconstructed DL CSIT, ${\bf \hat h}_{k}^{\sf dl}$ and ${\bf \Phi}_{k}$, the exact outer-product ${\bf h}_{k}^{\sf dl}\left({\bf h}_{k}^{\sf dl}\right)^{\sf H}$ is approximated by ${\bf\hat h}_k^{\sf dl}\left({\bf\hat h}_k^{\sf dl}\right)^{\sf H}+{\bf \Phi}_k$ with error matrix ${\bf \Delta}_k={\bf h}_k^{\sf dl}\left({\bf h}_k^{\sf dl}\right)^{\sf H}
        -\left({\bf\hat h}_k^{\sf dl}\left({\bf\hat h}_k^{\sf dl}\right)^{\sf H}+{\bf \Phi}_k\right)$. Then, the asymptotic error in the number of antennas is 
    \begin{align}
        &\lim_{N \to \infty} 
        \frac{1}{N^2}\left\|{\bf \Delta}_k\right\|_F^2 =\sum_{\ell\ne\ell'}2(1+{\rm Re}\{\eta\}^4)\left|{b}_{k,\ell}{b}_{k,\ell'}\right|^2-\sum_{\ell\ne\ell'}4{\rm Re}\{\eta\}^2{\rm Re}\left\{{g}_{k,\ell}^{\sf dl}\left({g}_{k,\ell'}^{\sf dl}\right)^*{g}_{k,\ell}^{\sf ul}\left({g}_{k,\ell'}^{\sf ul}\right)^*\right\}.
    \end{align}
\end{thm}
\begin{IEEEproof}
    See Appendix \ref{Proof_for_CovConvergence}.
\end{IEEEproof}

We provide our intuition for the proposed approximation in Theorem \ref{thm3}. The outer product of the DL channel is identical to our approximation in the average sense, i.e.,
\begin{align}
    \mathbb{E}\left[{\bf h}_k^{\sf dl}\left({\bf h}_k^{\sf dl}\right)^{\sf H}\right]=\mathbb{E}\left[\hat{{\bf h}}_k^{\sf dl}\left(\hat{{\bf h}}_k^{\sf dl}\right)^{\sf H}+{\bf\Phi}_k\right],\quad\forall k\in[K].
\end{align}
This claim is true because
\begin{align}
    &\mathbb{E}\left[\mathbb{E}\left[\left.{\bf h}_k^{\sf dl}\left({\bf h}_k^{\sf dl}\right)^{\sf H}-\hat{{\bf h}}_k^{\sf dl}\left(\hat{{\bf h}}_k^{\sf dl}\right)^{\sf H}~\right|~{\bf h}_k^{\sf ul},{\bf\Theta}_k,{\bf \Sigma}_k\right]\right]\nonumber\\
    &=\mathbb{E}_{{\bf h}_k^{\sf ul},{\bf\Theta}_k,{\bf \Sigma}_k}\left[\left(1-{\rm Re}\{\eta\}^2\right){\bf A}_k^{\sf dl}{\bf \Sigma}_k\left({\bf A}_k^{\sf dl}\right)^{\sf H}\right]\nonumber\\
    &=\mathbb{E}_{{\bf h}_k^{\sf ul},{\bf\Theta}_k,{\bf \Sigma}_k}\left[ {\bf \Phi}_k\right].
\end{align}
Consequently, our approximation is an unbiased estimator for ${\bf h}_k^{\sf dl}\left({\bf h}_k^{\sf dl}\right)^{\sf H}$. Furthermore, for the single-path scenario, our approximation becomes exact  asymptotically, which is stated in the following corollary.
\begin{cor}\label{cor:Outer_Product_Convergence_LOS}
    Under the LOS scenario, i.e., $L=1$, our approximation is tight, namely, 
    \begin{align}
        \lim_{N \to \infty}  \frac{1}{N^2}\left\|{\bf \Delta}_k\right\|_F^2 =0.
        \label{eq:Outer_Product_Convergence_LOS}
    \end{align}
\end{cor}
\begin{IEEEproof}
The proof is direct from the definition.
\end{IEEEproof}
 
\begin{figure}[t]
	\centering
    \includegraphics[width=0.5\linewidth]{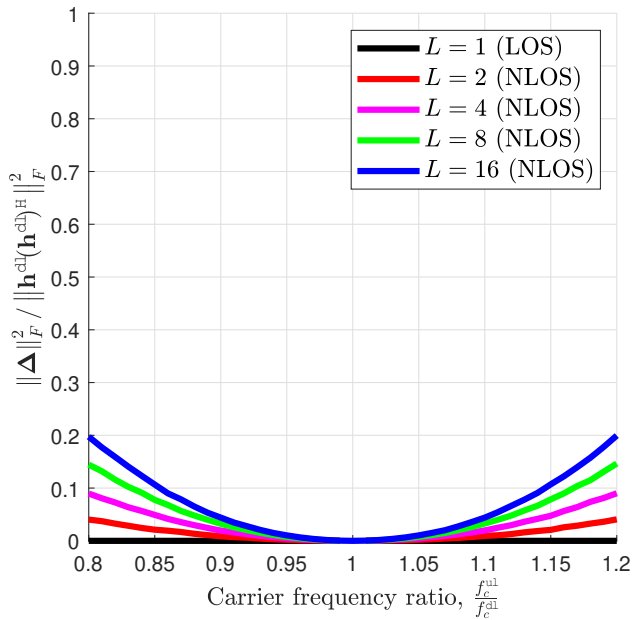}
  \caption{The approximation error as function of the ratio of UL and DL channels according to the number of channel paths $L\in \{1,2,4,8,16\}$.} \label{fig_FrobeniusNormAnalysis}
\end{figure}
To validate the CSIT approximation accuracy, we plot the normalized error as the carrier frequency ratio changes. Fig. \ref{fig_FrobeniusNormAnalysis} shows that with increasing number of channel paths, the normalized error increases, meaning the accuracy of our DL channel reconstruction method decreases. However, as long as the UL and DL frequency difference is small, the error remains negligible, indicating accurate estimation of DL channel outer product with minimal error.

From Theorem \ref{thm3}, we reformulate the DL sum-spectral efficiency in \eqref{eq:sumrate} as 
\begin{align}
   {\hat R}({\bf f}_1,\ldots, {\bf f}_K) 
    &=\log_2\left( \prod_{k=1}^{K}\frac{\sum_{i=1}^{K}{\bf f}_{i}^{\sf H} \left(\hat{{\bf h}}_k^{\sf dl}\left(\hat{{\bf h}}_k^{\sf dl}\right)^{\sf H}+{\bf\Phi}_k \right){\bf f}_{i}+\frac{{\sigma}_k^2}{P}}{\sum_{i\ne k}^{K}{\bf f}_{i}^{\sf H}\left(\hat{{\bf h}}_k^{\sf dl}\left(\hat{{\bf h}}_k^{\sf dl}\right)^{\sf H}+{\bf\Phi}_k\right){\bf f}_{i}+\frac{{\sigma}_k^2}{P}}\right), \label{eq:sumrate2}
\end{align}
where $ {\hat{\bf h}}_{k}^{\sf dl} ={\rm Re}\{{\eta}\}{\bf A}_k^{\sf dl}\left({\bf A}_k^{\sf ul}\right)^{\dagger} {\bf h}_k^{\sf ul}$ represents the reconstructed DL channel, and ${\bf \Phi}_k=\left(1-{\rm Re}\{{\eta}\}^2\right)\cdot$ ${\bf A}_k{\bf \Sigma}_k{\bf A}_k^{\sf H}$ is the error covariance matrix. The DL sum-spectral efficiency in \eqref{eq:sumrate2} is an approximation that takes into account imperfect channel state information at the BS. The BS optimizes this rate function for the DL transmission.


 The sum-spectral efficiency maximization is a well-known NP-hard problem \cite{luo2008dynamic}. The weighted MMSE algorithm is a widely used approach to tackle this optimization \cite{christensen2008weighted}. Recently, the GPIP algorithm was introduced, offering a comprehensive solution for joint user-selection, beamforming, and power allocation, and has shown to be the most effective solution for sum-spectral efficiency maximization in MU-MIMO systems, regardless of the number of antennas $N$ and users $K$ \cite{choi2019joint,han2020distributed,han2021sparse}. We take this approach for the robust DL precoding.

\subsection{Robust Precoding via GPIP }

The key idea of GPIP is to joinly optimize the precoding vectors $\left\{{\bf f}_1,\ldots, {\bf f}_K\right\}$. To accomplish this, we concatenate them into a high-dimensional optimization variable, i.e.,  
\begin{align}
     {\bf{f}} = [{\bf{f}}_{1}^{\sf{H}} ,\cdots, {\bf{f}}_{k}^{\sf{H}} ,\cdots, {\bf{f}}_{K}^{\sf{H}} ]^{\sf {H}}\in \mathbb{C}^{NK\times 1}. \label{eq:Large_BF}
\end{align} 
By utilizing the high-dimensional optimization variable, the expression of the sum-spectral efficiency in \eqref{eq:sumrate2} can be reformulated as a product of Rayleigh quotients, thus providing a more compact representation:
\begin{subequations}
\label{eq:P_Proposed_PCA}
\begin{align}
    \mathscr{P}^2:~&{\underset{{\bf f}\in \mathbb{C}^{NK\times 1}}{\text{arg~max}}}~ \prod_{k=1}^{K}\frac{{{\bf f}^{\sf H}{\bf A}_{k}{\bf f}}}{{{\bf f}^{\sf H}{\bf  B}_{k}{\bf f}}},\label{eq:P_Proposed_PCA_1}\\
    &\text{subject to } \|{\bf f}\|_2^2=1,\label{eq:P_Proposed_PCA_2}
\end{align}
\end{subequations}
where ${\bf A}_k\in\mathbb{C}^{NK\times NK}$ and ${\bf \Sigma}_k\in\mathbb{C}^{NK\times NK}$ are positive semi-definite block diagonal matrices defined as
\begin{align}
{\bf  A}_{k}&={\bf I}_K \otimes \left({\bf \hat h}_k^{\sf dl}\left({\bf \hat h}_k^{\sf dl}\right)^{\sf H}+{\bf \Phi}_k\right) + \frac{\sigma_k^2}{P}{\bf I}_{NK},\label{eq:A}\\
{\bf  B}_{k}&={\bf  A}_{k}-{\bf 1}_k{\bf 1}_k^{\sf T}\otimes \left({\bf \hat h}_k^{\sf dl}\left({\bf \hat h}_k^{\sf dl}\right)^{\sf H}+{\bf \Phi}_k\right). \label{eq:B}
\end{align}
As shown in \cite{choi2019joint}, the objective function in \eqref{eq:P_Proposed_PCA_1} is scale invariant, we can drop the power constraint in \eqref{eq:P_Proposed_PCA_2}, which boils down to 
\begin{align}
    \mathscr{P}^3:~{\underset{{\bf f}\in \mathbb{C}^{NK\times 1}}{\text{arg~max}}}~\prod_{k=1}^{K}\frac{{{\bf f}^{\sf H}{\bf A}_{k}{\bf f}}}{{{\bf f}^{\sf H}{\bf  B}_{k}{\bf f}}}.\label{eq:P_Proposed_Final}
\end{align}

Extending the result in \cite{choi2019joint}, we identify the local optimal solution for the sum-spectral efficiency maximization problem in \eqref{eq:P_Proposed_Final}. The following theorems state the first- and second-order optimality conditions. 

\begin{thm}\label{thm4}
Let $\gamma({\bf f})=\prod_{k=1}^{K}\frac{{{\bf f}^{\sf H}{\bf A}_{k}{\bf f}}}{{{\bf f}^{\sf H}{\bf  B}_{k}{\bf f}}}$. A stationary point ${\bf f}\in \mathbb{C}^{NK\times 1}$ for problem \eqref{eq:P_Proposed_Final} is an eigenvector of the following functional generalized eigenvalue problem:
\begin{align}
    {\bf \bar A}\left({\bf f}\right){\bf f}
    =\gamma({\bf f}){\bf \bar B}\left({\bf f}\right){\bf f},\label{eq:Generalized_Eigenvalue_Problem}
\end{align}
where
\begin{align}
    {\bf \bar A}\left({\bf f}\right) = \sum_{i=1}^{K}
    \left(
    \prod_{k\ne i}{\bf f}^{\sf H}{\bf A}_{k}{\bf f}
    \right){\bf A}_{i}
    ~and~{\bf \bar B}\left({\bf f}\right) = \sum_{i=1}^{K}
    \left(
    \prod_{k\ne i}{\bf f}^{\sf H}{\bf B}_{k}{\bf f}
    \right){\bf B}_{i}.
    \end{align}
\end{thm}
\begin{IEEEproof}
See Appendix \ref{Proof_for_LOC_First}.
\end{IEEEproof}
Theorem 4 shows that the stationary point of optimization problem in \eqref{eq:P_Proposed_Final} can be found by identifying the ${\bf f}$ that satisfies condition in \eqref{eq:Generalized_Eigenvalue_Problem}. This condition can be represented as a generalized eigenvalue problem: $\left[{\bf \bar B}\left({\bf f}\right)\right]^{-1}{\bf \bar A}\left({\bf f}\right){\bf f}=\gamma({\bf f}){\bf f}$, where $\gamma({\bf f})$ is the eigenvalue of $\left[{\bf \bar B}\left({\bf f}\right)\right]^{-1}{\bf \bar A}\left({\bf f}\right)$ and ${\bf f}$ is the corresponding eigenvector. The eigenvalue $\gamma({\bf f})$ also represents the objective function in \eqref{eq:P_Proposed_Final}. To maximize this function, we must find the first eigenvector of $\left[{\bf \bar B}\left({\bf f}\right)\right]^{-1}{\bf \bar A}\left({\bf f}\right)$.

\begin{thm} \label{thm5}
Let ${\bf f}^{\star}$ be the solution of Theorem \ref{thm4}. This stationary point ${\bf f}^{\star}$ is a local-optimum, provided that 
\begin{align}
    &\rho_{\sf min}\left(\sum_{i=1}^{K}   \frac{{\bf A}_i^{\sf H}{\bf f}^{\star}({\bf f}^{\star})^{\sf H}{\bf A}_i}{\left(({\bf f}^{\star})^{\sf H}{\bf A}_i{\bf f}^{\star}\right)^2} \right) >
    \rho_{\sf max}\left(\sum_{i=1}^{K}   \frac{{\bf B}_i^{\sf H}{\bf f}^{\star}({\bf f}^{\star})^{\sf H}{\bf B}_i}{\left(({\bf f}^{\star})^{\sf H}{\bf B}_i{\bf f}^{\star}\right)^2}\right).    \label{eq:Sub_Opt_Condition}
\end{align}
\end{thm}
 \begin{IEEEproof}
See Appendix \ref{Proof_for_LOC_Second}.
\end{IEEEproof}
According to Theorem \ref{thm5}, if the minimum eigenvalue of $\sum_{i=1}^{K} \frac{{\bf A}_i^{\sf H}{\bf f}^{\star}({\bf f}^{\star})^{\sf H}{\bf A}_i}{\left(({\bf f}^{\star})^{\sf H}{\bf A}i{\bf f}^{\star}\right)^2}$ is greater than the maximum eigenvalue of $\sum_{i=1}^{K} \frac{{\bf B}_i^{\sf H}{\bf f}^{\star}({\bf f}^{\star})^{\sf H}{\bf B}_i}{\left(({\bf f}^{\star})^{\sf H}{\bf B}_i{\bf f}^{\star}\right)^2}$, then the stationary point ${\bf f}^{\star}$ has a direction of strictly negative curvature. This eigenvalue test allows us to determine if a stationary point ${\bf f}^{\star}$ is the local optimal solution for a non-convex optimization problem. The maximum and minimum eigenvalues can be calculated using either power iteration or inverse power iteration algorithms.

\begin{algorithm}[t]
\caption{Proposed Robust DL Precoding}
Initialization: $t=1,~ {\bf f}^{(1)}={\sf ZF},~ {\bf f}^{(0)}={\bf 0},~{\text{and }} \epsilon$\;
    \While{$t\ne 0$}
    {   $t\gets t+1$;\;
    \eIf{$\left|\gamma\left({\bf f}^{(t-1)}\right)-\gamma\left({\bf f}^{(t)}\right)\right|/\gamma\left({\bf f}^{(t-1)}\right)\ge\epsilon$}
    {
        ${\bf f}^{(t)} \gets \left[{\bf \bar B}\left({\bf f}^{(t-1)}\right)\right]^{-1}{\bf \bar A}\left({\bf f}^{(t-1)}\right){\bf f}^{(t-1)}$;\;
        ${\bf f}^{(t)} \gets \frac{{\bf f}^{(t)}}{\left\|{\bf f}^{(t)}\right\|_2}$;\;
    }{
        \eIf{Theorem \ref{thm5} holds}
        {
            Break;
        }{
            $t=1 \text{ and }{\bf f}^{(1)} = $ Random beamforming;\;
            Continue;
        }
    }
}
\label{alg:Proposed}
\end{algorithm}

We present a computationally efficient algorithm that identifies the solution satisfying the first and second order optimality conditions derived in Theorem \ref{thm4} and \ref{thm5}, as introduced in our companion paper \cite{choi2019joint,han2020distributed,han2021sparse}. The proposed algorithm iteratively finds the local-optimum ${\bf f}^{\star}$. At each iteration, it starts by constructing the functional matrices ${\bf \bar A}\left({\bf f}^{(t-1)}\right)$ and $ {\bf \bar B}\left({\bf f}^{(t-1)}\right)$ based on the precoding vector ${\bf f}^{(t-1)}$ obtained in the previous iteration. Next, the first eigenvector of $\left[ {\bf \bar B}\left({\bf f}^{(t-1)}\right) \right]^{-1}
{\bf \bar A}\left({\bf f}^{(t-1)}\right)$ is found through a power-iteration process \cite{lee2008achievable}. The updated ${\bf f}^{(t)}$ is then obtained using the equation: ${\bf f}^{(t)}=\left[ {\bf \bar B}\left({\bf f}^{(t-1)}\right) \right]^{-1}
{\bf \bar A}\left({\bf f}^{(t-1)}\right){\bf f}^{(t-1)}$ and normalized to have a unit length.
The iteration continues until the objective function converges, with the stopping criteria being $\left|\gamma\left({\bf f}^{(t-1)}\right)-\gamma\left({\bf f}^{(t)}\right)\right|/\gamma\left({\bf f}^{(t-1)}\right)\le\epsilon$, where $\epsilon$ is a small positive value. After the algorithm reaches convergence at $t=T$, the solution ${\bf f}^{(T)}$ is checked against the second-order necessary optimality condition of Theorem \ref{thm5}. If it satisfies the condition, the algorithm ends with ${\bf f}^{(T)}$. Otherwise, it restarts with a new starting vector. The entire process is summarized in Algorithm \ref{alg:Proposed}.

\begin {table}[t]
\footnotesize
\caption {System-Level Simulation Parameters} \vspace{-0.1cm}\label{tab:Sys_Assumption} 
  	 \begin{center}
  \begin{tabular}{ l  c }
    \hline\hline
    Parameters & Value  \\ \hline
            BS topology & Single hexagonal cell with ISD 500m \\

        User distribution & Uniform per cell  \\ 
        UL carrier frequency & 10 GHz \\
        DL carrier frequency & 12 GHz ($f_c^{\sf dl}/f_c^{\sf ul}=1.2$)\\
        The number of users $K$ & $16$\\
        Noise power & -113dB\\
        Path-loss model & Standard model at TR 38.901\\
        BS/UE height & 32m/1.5m\\ \hline
  \end{tabular}
\end{center}\vspace{-0.3cm}
\end {table}

\begin{rem}[Computational complexity of the proposed precoding framework]
The computational complexity of our precoding algorithm is of the order of $\mathcal{O}(JN^2K)$, as reported in various studies including \cite{choi2019joint,han2020distributed,han2021sparse,park2022rate}. This is a significant improvement over the WMMSE method \cite{christensen2008weighted} whose complexity order is $\mathcal{O}\left(J(KN)^{3.5}\right)$. Our algorithm demonstrates more computational efficiency compared to the WMMSE approach.
\end{rem}

\section{Simulations Results}\label{sec:Simulation_Results}
In this section, we present a comprehensive comparison of the ergodic sum-spectral efficiency achieved by our proposed algorithm with existing precoding schemes \cite{christensen2008weighted} using system-level simulations. The simulation parameters and network topology are outlined in detail in Table \ref{tab:Sys_Assumption}. Our simulations consider a fixed base station location, with randomly distributed user locations for each scenario. This ensures a fair evaluation of the algorithm's performance under various network conditions.

\begin{figure}[t]
	\centering
    \includegraphics[width=0.5\linewidth]{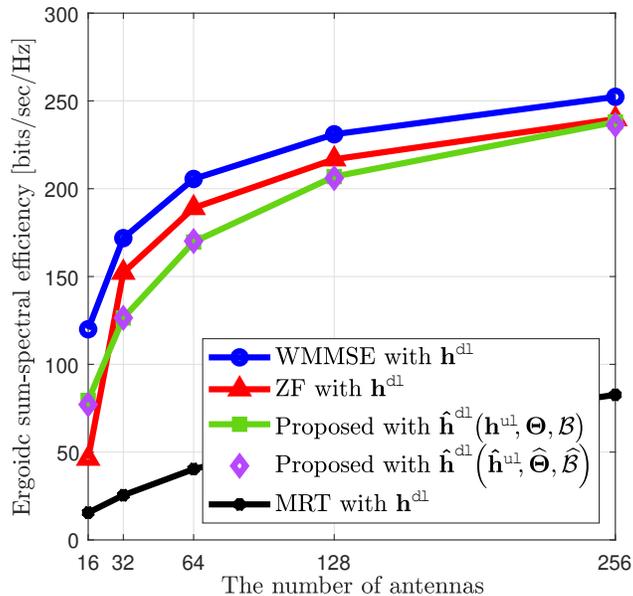}
  \caption{The ergodic sum-spectral efficiency with $(L,K)=(3,16)$ as increasing the number of antennas.} \label{fig_K16R250L3UpperMid}
\end{figure}

\begin{figure}[t]
	\centering
    \includegraphics[width=0.5\linewidth]{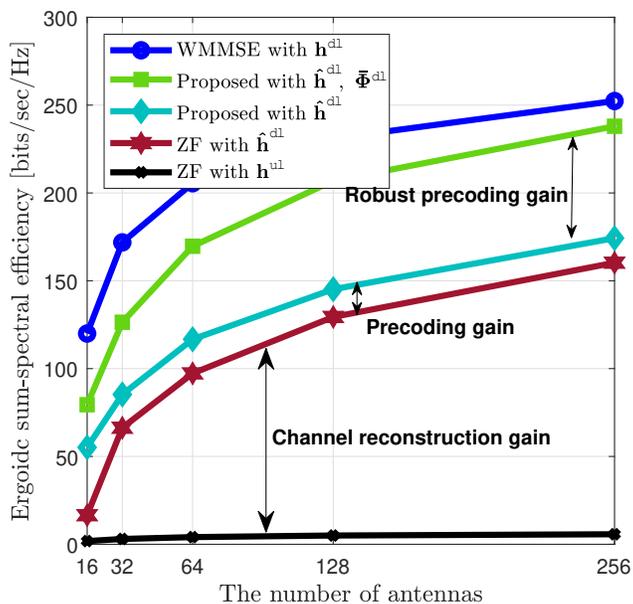}
  \caption{The ergodic sum-spectral efficiency with $(L,K)=(3,16)$ according to channel knowledge and precoding strategies. } \label{fig_K16R250L3UpperMid_GainAnalysis}
\end{figure}

\subsection{Effects of Geometric Parameter Estimation Error}

Fig. \ref{fig_K16R250L3UpperMid} shows the growth of the ergodic sum-spectral efficiency as the number of antennas increases. The proposed algorithm is evaluated in two scenarios: with perfect knowledge of the geometry parameters $\left\{{\bf h}_k^{\sf ul},{\bf\Theta}_k,{\bf \Sigma}_k\right\}$ and with imperfect knowledge estimated via spatial smoothing and least-squared estimation $\left\{{\bf \hat h}_k^{\sf ul},{\bf \hat \Theta}_k,{\bf \hat \Sigma}_k\right\}$. The results show that the proposed algorithm attains the ergodic sum-spectral efficiency of ZF and WMMSE when evaluated with perfect DL CSIT. When evaluated with estimated geometry parameters, the degradation caused by estimation error is minimal.

\subsection{Effects of CSIT Knowledge and Precoding}
Fig. \ref{fig_K16R250L3UpperMid_GainAnalysis} compares the achievable ergodic sum-spectral efficiency gains based on CSIT knowledge and precoding strategies. ZF precoding with perfect UL channel knowledge is plotted as a benchmark, followed by ZF precoding with DL channel reconstructed from UL channel. The improvement in performance is significant, demonstrating the effectiveness of the DL reconstruction method. The proposed GPI precoding outperforms ZF precoding when both uses the DL reconstructed channel knowledge, due to precoding optimization. When using both DL channel knowledge and DL channel covariance, the proposed GPIP improves the sum-spectral efficiency greatly and the performance gap with WMMSE decreases with increasing antenna numbers.

\begin{figure}[t]
	\centering
    \includegraphics[width=0.5\linewidth]{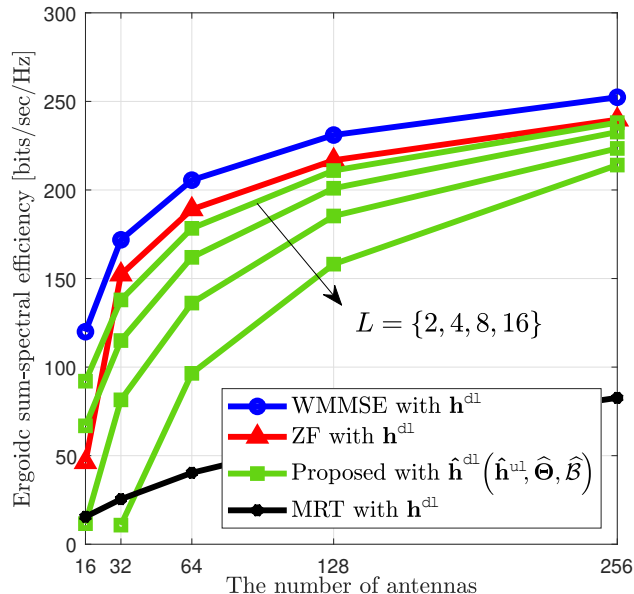}
  \caption{The ergodic sum-spectral efficiency as increasing the number of channel paths $L$. The number of users is set to be $K=16$.} \label{fig_K16R250L24816UpperMid}
\end{figure}

 \subsection{Effects of the Number of Channel Paths}

Fig. \ref{fig_K16R250L24816UpperMid} demonstrates the deterioration of ergodic sum-spectral efficiency as the number of paths increases. The simulations were conducted with $L$ values of 2, 4, 8, and 16. As the number of paths increases, two algorithmic issues arise: 1) an increase in the estimation error of geometric parameters, leading to a decline in DL channel reconstruction, and 2) a reduction in orthogonality in the spatial domain, resulting in a decrease in robust precoding gain. These simulation results highlight the robustness of the proposed framework to both channel reconstruction error and moderate values of $N/L$. As seen in Fig. \ref{fig_K16R250L24816UpperMid}, the proposed algorithm remains robust even when $N/L$ exceeds 16, which is equivalent to 256 divided by 16.

\begin{figure}[t]
    \centering
    \includegraphics[width=0.5\linewidth]{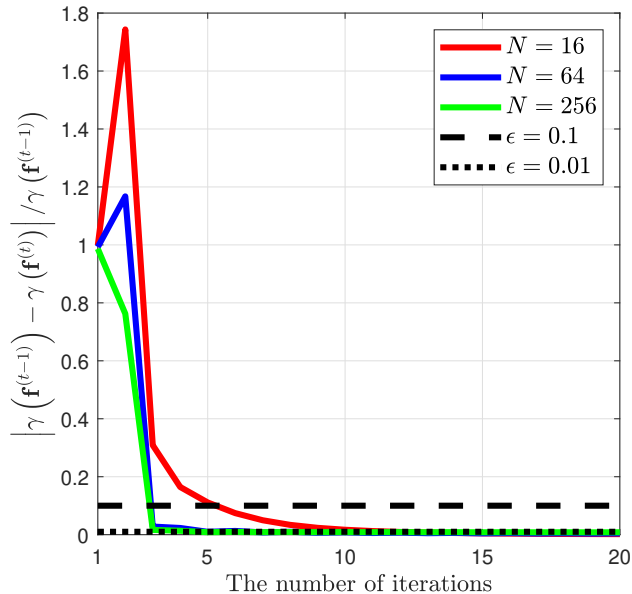}
    \caption{The convergence rate of the proposed GPIP when $\epsilon=0.1$ and $\epsilon=0.01$.} \label{fig_CovergenceSpeed}
\end{figure}

\subsection{Convergence Property}
Fig. \ref{fig_CovergenceSpeed} illustrates the convergence rate of Algorithm \ref{alg:Proposed}. Our investigation focuses on determining the number of iterations required to meet the stopping criteria as we increase the number of BS antennas, i.e., $N\in\left\{16,64,256\right\}$. As depicted in Fig. \ref{fig_CovergenceSpeed}, the proposed algorithm requires only five iterations to reach the desired precision of $\epsilon=0.1$. The maximum number of iterations necessary to achieve $\epsilon=0.01$ is ten, indicating a fast convergence rate for the proposed algorithm.

\section{Conclusion}
We introduced a novel robust DL data transmission framework that provides substantial sum-spectral efficiency gains in FDD massive MIMO systems without CSI feedback. Our approach leverages the geometric  reciprocity between UL and DL channels to reconstruct the DL channel. Using these reconstructed DL channel and the corresponding error covariance matrix, we developed a robust DL precoding method that maximizes sum-spectral efficiency. Our major finding was that the FDD massive MIMO gains is still attainable without CSI training and feedback. Throughout the system-level simulations, we verified our results.



\appendix

\subsection{Proof for Lemma \ref{lem1}}\label{Proof_for_Corr_UL_DL}
Let us assume that the phase of path attenuation is uniformly distributed, i.e., $\angle{g_{k,\ell}^{\sf ul}} =\frac{2\pi}{\lambda^{\sf ul}}r_{k,\ell}-\phi_{k,\ell} \in\left[0,2\pi\right)$. Then, 
\begin{align}
     \left[\mathbb{E}[{\bf g }_k^{\sf dl}({\bf g }_k^{\sf ul})^{\sf H}]\right]_{\ell,\ell} 
    &=\frac{|g_{k,\ell}|^2}{2\pi}\int_0^{2\pi} e^{-j(\frac{\lambda^{\sf ul}}{\lambda^{\sf dl}}-1)\angle{\frac{2\pi}{\lambda^{\sf ul}}r_k}}\, d\angle{{2\pi}/{\lambda^{\sf ul}}r_k}\\
    &=|g_{k,\ell}|^2\frac{\lambda^{\sf dl}}{2\pi(\lambda^{\sf ul}-\lambda^{\sf dl})}\left( \sin\left(2\pi\frac{\lambda^{\sf ul}}{\lambda^{\sf dl}}\right)-2j\sin^2\left(\pi\frac{\lambda^{\sf ul}}{\lambda^{\sf dl}}\right)\right).
\end{align}
In addition, with $\ell\ne\ell'$,
$\left[\mathbb{E}[{\bf g }_k^{\sf dl}({\bf g }_k^{\sf ul})^{\sf H}]\right]_{\ell,\ell'}=0$ provided that the paths are uncorrelated. This completes the proof.

\subsection{Proof for Theorem \ref{thm:OptimalDlChannelEstimator}}\label{Proof_for_OptimalDlChannelEstimator}
The optimal estimator can be obtained by calculating following conditional mean, i.e., 
\begin{align}
    {\bf\hat h}_k^{\sf dl, optimal} 
    &= \mathbb{E}\left[\left.{\bf h}_k^{\sf dl}~\right|~{\bf h}_k^{\sf ul},{\bf\Theta}_k,{\bf \Sigma}_k\right]\nonumber\\
    &={\bf A}_k^{\sf dl}\mathbb{E}\left[\left.{\bf g}_k^{\sf dl}~\right|~{\bf h}_k^{\sf ul},{\bf\Theta}_k,{\bf \Sigma}_k\right].\label{eq:proof1}
\end{align}
To ease of expansion, we denote the phase of ${\bf g}_k^{\sf dl}$ as $e^{j\angle{\bf g}_k^{\sf dl}}=\left[e^{j\angle{g_{k,1}^{\sf dl}}},\cdots,e^{j\angle{g_{k,L}^{\sf dl}}}\right]^{\sf T}$ in a element-wise manner and the path attenuation as $\left|{\bf g}_k^{\sf dl}\right|=\left[\left|{g_{k,1}}\right|,\cdots,\left|{\beta_{k,L}}\right|\right]^{\sf T}$ in a similar manner so that ${\bf g}_k^{\sf dl}=\left|{\bf g}_k^{\sf dl}\right|\odot e^{j\angle{\bf g}_k^{\sf dl}}$. In addition, the expectation in \eqref{eq:proof1} is taken over the phase term, i.e., 
\begin{align}
     \mathbb{E}\left[\left.{\bf g}_k^{\sf dl}~\right|~{\bf h}_k^{\sf ul},{\bf\Theta}_k,{\bf \Sigma}_k\right] 
    &= \left|{\bf g}_k^{\sf dl}\right|\odot
    \int e^{j\angle{\bf g}_k^{\sf dl}}f\left(\left.\angle{\bf g}_k^{\sf dl}~\right|~{\bf h}_k^{\sf ul},{\bf\Theta}_k,{\bf \Sigma}_k\right)\, d\angle{\bf g}_k^{\sf dl}\nonumber\\
    &=\left|{\bf g}_k^{\sf dl}\right|\odot e^{j\frac{\lambda^{\sf ul}}{\lambda^{\sf dl}}\angle{\bf g}_k^{\sf ul}}\odot
    \int_0^{2\pi} e^{j\left(1-\frac{\lambda^{\sf ul}}{\lambda^{\sf dl}}\right){\boldsymbol\phi}_k}\frac{1}{2\pi}\, d{\boldsymbol\phi}_k\nonumber\\    
    &=\frac{\lambda^{\sf dl}}{2\pi(\lambda^{\sf ul}-\lambda^{\sf dl})}\left(\sin\left(2\pi\frac{\lambda^{\sf ul}}{\lambda^{\sf dl}}\right)-2j\sin^2\left(\pi\frac{\lambda^{\sf ul}}{\lambda^{\sf dl}}\right)\right)\cdot\left|{\bf g}_k^{\sf dl}\right|\odot e^{j\frac{\lambda^{\sf ul}}{\lambda^{\sf dl}}\angle{\bf g}_k^{\sf ul}}\nonumber\\
    &=\eta{\bf \Sigma}_k^{\frac{1}{2}}\left({\bf \Sigma}_k^{-\frac{1}{2}}\left({\bf A}_k^{\sf ul}\right)^{\dagger}{\bf h}_k^{\sf ul}\right)^{\frac{\lambda^{\sf ul}}{\lambda^{\sf dl}}}.
\end{align}
Therefore, by merging it with \eqref{eq:proof1}, we obtain the following optimal estimator, i.e., 
\begin{align}
    {\bf\hat h}_k^{\sf dl, optimal}
    &={\bf A}_k^{\sf dl}\mathbb{E}\left[\left.{\bf g}_k^{\sf dl}~\right|~{\bf h}_k^{\sf ul},{\bf\Theta}_k,{\bf \Sigma}_k\right]=\eta{\bf A}_k^{\sf dl}{\bf \Sigma}_k^{\frac{1}{2}}\left({\bf \Sigma}_k^{-\frac{1}{2}}\left({\bf A}_k^{\sf ul}\right)^{\dagger}{\bf h}_k^{\sf ul}\right)^{\frac{\lambda^{\sf ul}}{\lambda^{\sf dl}}}.
\end{align}
This completes proof.

\subsection{Proof for Corollary \ref{thm:LinearDlChannelEstimator}}\label{Proof_for_Corr1}
To derive the LMMSE DL channel estimate, we need to solve the following optimziation problem:
\begin{align}
    {\bf W}_k &= 
    {\underset{{\bf W}_k \in \mathbb{C}^{N\times N}}{\text{arg~min}}}~
    \mathbb{E}\left[\left.\left\|{\bf W}_k{\bf h}_k^{\sf ul}-{\bf h}_k^{\sf dl}\right\|_2^2~\right|~{\bf h}_k^{\sf ul},{\bf\Theta}_k,{\bf \Sigma}_k\right].
    \label{eq:MMSE}
\end{align}
Since the optimization problem in \eqref{eq:MMSE} is convex in ${\bf W}_k$, it is sufficient to find the KKT conditions to obtain the optimal solution. By taking the derivative with respect to ${\bf W}_k$, and setting it to zero, we attain 
    \begin{align}
 {\bf W}_{k} = \frac{\lambda^{\sf dl}}{2\pi(\lambda^{\sf ul}-\lambda^{\sf dl})}\sin\left(2\pi\frac{\lambda^{\sf ul}}{\lambda^{\sf dl}}\right){\bf A}_k^{\sf dl}\left({\bf A}_k^{\sf ul}\right)^{\dagger}.
    \end{align}
    Therefore, the DL channel estimates can be obtained by directly adopting the results in \eqref{eq:col1}, i.e., 
\begin{align}
    {\hat{\bf h}}_{k}^{\sf dl, L-MMSE} &= {\bf W}_k{\bf h}_k^{\sf ul}\nonumber\\
  & =\frac{\lambda^{\sf dl}}{2\pi(\lambda^{\sf ul}-\lambda^{\sf dl})}\sin\left(2\pi\frac{\lambda^{\sf ul}}{\lambda^{\sf dl}}\right){\bf A}_k^{\sf dl}{\bf g}_k^{\sf ul}.\label{eq:MMSE_Estimates}
\end{align}
This ends proof.
\subsection{Proof for Theorem \ref{thm:MseMatrxOfOptimal}}\label{ProofForThm2}
To ease of expansion, we denote that ${\bf\bar g}_k^{\sf ul}={\bf \Sigma}_k^{\frac{1}{2}}\left({\bf \Sigma}_k^{-\frac{1}{2}}\left({\bf A}_k^{\sf ul}\right)^{\dagger}{\bf h}_k^{\sf ul}\right)^{\frac{\lambda^{\sf ul}}{\lambda^{\sf dl}}}$, i.e., ${\bf\hat h}_k^{\sf dl, MMSE} = \eta{\bf A}_k^{\sf dl}{\bf\bar g}_k^{\sf ul}$ . Then, 
\begin{align}
    {\bf\Phi}_k^{\sf MMSE} &=\mathbb{E}\left[\left.\left({\bf h}_k^{\sf dl}-{\bf\hat h}_k^{\sf dl, MMSE}\right)\left({\bf h}_k^{\sf dl}-{\bf\hat h}_k^{\sf dl, MMSE}\right)^{\sf H}\right|{\bf h}_k^{\sf ul},{\bf\Theta}_k,{\bf \Sigma}_k\right]\nonumber\\
    &=\mathbb{E}\left[{\bf A}_k^{\sf dl}{\bf g}_k^{\sf ul}\left({\bf g}_k^{\sf ul}\right)^{\sf H}\left({\bf A}_k^{\sf dl}\right)^{\sf H}+|\eta|^2{\bf A}_k^{\sf dl}{\bf\bar g}_k^{\sf ul}\left({\bf\bar g}_k^{\sf ul}\right)^{\sf H}\left({\bf A}_k^{\sf dl}\right)^{\sf H}\right.\nonumber\\
    &\left.\quad\quad\quad\quad\quad\quad\quad\quad\quad\quad\quad\quad-\eta^*{\bf A}_k^{\sf dl}{\bf g}_k^{\sf ul}\left({\bf\bar g}_k^{\sf ul}\right)^{\sf H}\left({\bf A}_k^{\sf dl}\right)^{\sf H}-\eta{\bf A}_k^{\sf dl}{\bf\bar g}_k^{\sf ul}\left({\bf g}_k^{\sf ul}\right)^{\sf H}\left({\bf A}_k^{\sf dl}\right)^{\sf H}\right]\nonumber\\
    &=\mathbb{E}\left[{\bf A}_k^{\sf dl}{\bf\Sigma}_k\left({\bf A}_k^{\sf dl}\right)^{\sf H}+|\eta|^2{\bf A}_k^{\sf dl}{\bf\Sigma}_k\left({\bf A}_k^{\sf dl}\right)^{\sf H}-\eta^*\eta{\bf A}_k^{\sf dl}{\bf\Sigma}_k\left({\bf A}_k^{\sf dl}\right)^{\sf H}-\eta\eta^*{\bf A}_k^{\sf dl}{\bf\Sigma}_k\left({\bf A}_k^{\sf dl}\right)^{\sf H}\right]\label{eq:ProofForThm2}\\
    &=\left(1-|\eta|^2\right){\bf A}_k^{\sf dl}{\bf \Sigma}_k\left({\bf A}_k^{\sf dl}\right)^{\sf H},    
\end{align}
where the equality in \eqref{eq:ProofForThm2} follows from Lemma \ref{lem1}. This completes the proof.


\subsection{Proof for Corollary \ref{cor:MMSE_Derivation}}\label{Proof_for_Cor2}
The MSE of each estimator can be obtained by using MSE matrix, i.e., 
\begin{align}
    {\sf MSE} 
    &=\lim_{N\rightarrow \infty}\frac{1}{N} {\sf Tr}\left\{
    \left(1-|\eta|^2\right){\bf A}_k^{\sf dl}{\bf \Sigma}_k\left({\bf A}_k^{\sf dl}\right)^{\sf H}
    \right\} / {\sf Tr}\left\{{\bf \Sigma}_k\right\} \nonumber\\
    &=\lim_{N\rightarrow \infty}\frac{1}{N} {\sf Tr}\left\{
    \left(1-|\eta|^2\right){\bf \Sigma}_k\left({\bf A}_k^{\sf dl}\right)^{\sf H}{\bf A}_k^{\sf dl}
    \right\} / {\sf Tr}\left\{{\bf \Sigma}_k\right\}= \left(1-|\eta|^2\right){\sf Tr}\left\{
    {\bf \Sigma}_k
    \right\} / {\sf Tr}\left\{{\bf \Sigma}_k\right\}=1-|\eta|^2,\nonumber
\end{align}
where the third equality follows that $\lim_{N\rightarrow \infty}\frac{1}{N} {\sf Tr}\left\{
    \left({\bf A}_k^{\sf dl}\right)^{\sf H}{\bf A}_k^{\sf dl}
    \right\}={\bf I}$. Similarly, ${\sf L\text{-}MSE}$ is obtained as ${\sf L\text{-}MSE} 
    = \lim_{N\rightarrow \infty}\frac{1}{N}{\sf Tr}\left\{{\bf \Phi}_k^{\sf L-MMSE}\right\}/ [{\sf Tr}\left\{{\bf \Sigma}_k\right\}=1-{\rm Re}\{\eta\}^2.$
This ends proof.




\subsection{Proof for Theorem \ref{thm3}}\label{Proof_for_CovConvergence}
We commence by representing the Frobenius norm by a trace operation, i.e., $\|{\bf A}\|_F^2={\sf Tr}\{{\bf A}{\bf A}^{\sf H}\}$. 
\begin{align}
    &\left\|{\bf h}_k^{\sf dl}\left({\bf h}_k^{\sf dl}\right)^{\sf H}
    -\left(
    {\bf\hat h}_k^{\sf dl}\left({\bf\hat h}_k^{\sf dl}\right)^{\sf H}+{\bf\Phi}_k
    \right)\right\|_F^2\nonumber\\
    &=\left\|
    {\bf A}_k^{\sf dl}\left({\bf g}_k^{\sf dl}\left({\bf g}_k^{\sf dl}\right)^{\sf H}
    -\left({\rm Re}\{\eta\}^2{\bf g}_k^{\sf ul}\left({\bf g}_k^{\sf ul}\right)^{\sf H}+\left(1-{\rm Re}\{\eta\}^2\right){\bf \Sigma}_k
    \right)
    \right)\left({\bf A}_k^{\sf dl}\right)^{\sf H}
    \right\|_F^2.
\end{align}
To ease of expansion, we denote that ${\bf M}_k={\bf g}_k^{\sf dl}\left({\bf g}_k^{\sf dl}\right)^{\sf H}
    -\left({\rm Re}\{\eta\}^2{\bf g}_k^{\sf ul}\left({\bf g}_k^{\sf ul}\right)^{\sf H}
    +\left(1-{\rm Re}\{\eta\}^2\right){\bf \Sigma}_k
    \right)$. Then,
\begin{align}
    \left\|
    {\bf A}_k^{\sf dl}{\bf M}_k\left({\bf A}_k^{\sf dl}\right)^{\sf H}
    \right\|_F^2&={\sf Tr}
    \left\{
    {\bf A}_k^{\sf dl}{\bf M}_k\left({\bf A}_k^{\sf dl}\right)^{\sf H}
    {\bf A}_k^{\sf dl}{\bf M}_k^{\sf H}\left({\bf A}_k^{\sf dl}\right)^{\sf H}
    \right\}={\sf Tr}
    \left\{
    \left({\bf A}_k^{\sf dl}\right)^{\sf H}{\bf A}_k^{\sf dl}{\bf M}_k\left({\bf A}_k^{\sf dl}\right)^{\sf H}
    {\bf A}_k^{\sf dl}{\bf M}_k^{\sf H}
    \right\}.\label{eq:Thm3-1}
\end{align}
Since $\lim_{N \to \infty} \frac{1}{N}\left({\bf A}_k^{\sf dl}\right)^{\sf H}
    {\bf A}_k^{\sf dl} = {\bf I}_N $, we can simplify the term in \eqref{eq:Thm3-1} as follows:
\begin{align}
    &\lim_{N \to \infty} \frac{1}{N^2}{\sf Tr}
    \left\{
    \left({\bf A}_k^{\sf dl}\right)^{\sf H}{\bf A}_k^{\sf dl}{\bf M}_k\left({\bf A}_k^{\sf dl}\right)^{\sf H}
    {\bf A}_k^{\sf dl}{\bf M}_k^{\sf H}
    \right\}\nonumber\\
    &={\sf Tr}
    \left\{
    {\bf M}_k{\bf M}_k^{\sf H}
    \right\}
    ={\sf Tr}
    \left\{
    {\bf M}_k^2
    \right\}\nonumber\\
    &=\sum_{\ell\ne\ell'}2(1+{\rm Re}\{\eta\}^4)\left|{b}_{k,\ell}{b}_{k,\ell'}\right|^2-\sum_{\ell\ne\ell'}4{\rm Re}\{\eta\}^2{\rm Re}\left\{{g}_{k,\ell}^{\sf dl}\left({g}_{k,\ell'}^{\sf dl}\right)^*{g}_{k,\ell}^{\sf ul}\left({g}_{k,\ell'}^{\sf ul}\right)^*\right\}.
\end{align}
This ends proof.
\subsection{Proof for Theorem \ref{thm4}}\label{Proof_for_LOC_First}
We commence by defining the Lagrange function, i.e., $\mathcal{L}\left({\bf f}\right) = \prod_{k=1}^{K}
    \frac{{\bf f}^{\sf H}{\bf A}_{k}{\bf f}}
    {{\bf f}^{\sf H}{\bf B}_{k}{\bf f}}$. To find a stationary point, we calculate the partial derivatives of $\mathcal{L}({\bf f})$ with respect to ${\bf f}^{\sf H}$, and set them equal to zero. The derivative with respect to ${\bf f}^{\sf H}$ yields
\begin{align}
    &\nabla_{{\bf f}^{\sf H}}\left\{\mathcal{L({\bf f})}\right\}={\bf 0}\Leftrightarrow
    \sum_{i=1}^{K}
    \left(
    \prod_{k\ne i}{\bf f}^{\sf H}{\bf A}_{k}{\bf f}
    \right)\left(\prod_{k=1}^{K}{\bf f}^{\sf H}{\bf B}_{k}{\bf f}\right){\bf A}_{i}
    {\bf f}=\sum_{i=1}^{K}
    \left(
    \prod_{k\ne i}{\bf f}^{\sf H}{\bf B}_{k}{\bf f}
    \right)\left(\prod_{k=1}^{K}{\bf f}^{\sf H}{\bf A}_{k}{\bf f}\right){\bf B}_{i}
    {\bf f}. \label{eq:derivative_f}
   \end{align} 
   By rearranging \eqref{eq:derivative_f}, we obtain: $\sum_{i=1}^{K}
    \left(
    \prod_{k\ne i}{\bf f}^{\sf H}{\bf A}_{k}{\bf f}
    \right){\bf A}_{i}
    {\bf f}= \gamma({\bf f})\sum_{i=1}^{K}
    \left(
    \prod_{k\ne i}{\bf f}^{\sf H}{\bf B}_{k}{\bf f}
    \right){\bf B}_{i}
    {\bf f}. $ This simplifies to ${\bf \bar A}\left({\bf f}\right){\bf f}=
    \gamma({\bf f}){\bf \bar B}\left({\bf f}\right){\bf f}.$ This completes the proof.

\subsection{Proof for Theorem \ref{thm5}}\label{Proof_for_LOC_Second}
To show the local-optimality, it is sufficient to show that the Hessian matrix at a stationary point is negative definite. To examine this condition, we first derive the Hessian matrix evaluated at an arbitrary point ${\bf f}\in \mathbb{C}^{LNK\times 1}$, which is given by 
\begin{align}
    \nabla_{{\bf f}^{\sf H}}^2\mathcal{L}\left({\bf f}\right)=2\left\{\nabla_{{\bf f}^{\sf H}}\mathcal{L}\left({\bf f}\right)\right\}\left(
    \sum_{i=1}^{K}\frac{{\bf A}_i{\bf f}}{{\bf f}^{\sf H}{\bf A}_i{\bf f}}
    -\sum_{i=1}^{K}\frac{{\bf B}_i{\bf f}}{{\bf f}^{\sf H}{\bf B}_i{\bf f}}\right)^{\sf H}+2\mathcal{L}\left({\bf f}\right)\left\{\nabla_{{\bf f}^{\sf H}}\left(
    \sum_{i=1}^{K}\frac{{\bf A}_i{\bf f}}{{\bf f}^{\sf H}{\bf A}_i{\bf f}}
    -\sum_{i=1}^{K}\frac{{\bf B}_i{\bf f}}{{\bf f}^{\sf H}{\bf B}_i{\bf f}}\right)\right\}.\label{eq:Hessian_matrix_1}
\end{align}
By plugging a stationary point ${\bf f}^{\star}$ obtained from Theorem 3 into \eqref{eq:Hessian_matrix_1}, it follows that
\begin{align}
    \nabla_{{\bf f}^{\sf H}}^2\mathcal{L}({\bf f}^{\star})&=
    2\nabla_{{\bf f}^{\sf H}}\mathcal{L}\left({\bf f}^{\star}\right)\cdot
    \left(
    \sum_{i=1}^{K}\frac{{\bf A}_i{\bf f}^{\star}}{\left(({\bf f}^{\star})^{\sf H}{\bf A}_i{\bf f}^{\star}\right)}-\sum_{i=1}^{K}\frac{{\bf B}_i{\bf f}^{\star}}{\left(({\bf f}^{\star})^{\sf H}{\bf B}_i{\bf f}^{\star}\right)}\right)^{\sf H}\nonumber\\
    &+2\mathcal{L}({\bf f}^{\star})
    \left(
    \sum_{i=1}^{K}\frac{-2{\bf A}_i{\bf f}^{\star}\left({\bf f}^{\star}\right)^{\sf H}{\bf A}_i}{(\left({\bf f}^{\star}\right)^{\sf H}{\bf A}_i{\bf f}^{\star})^2}+\sum_{i=1}^{K}\frac{2{\bf B}_i{\bf f}^{\star}\left({\bf f}^{\star}\right)^{\sf H}{\bf B}_i}{(\left({\bf f}^{\star}\right)^{\sf H}{\bf B}_i{\bf f}^{\star})^2}
    \right).
    \label{eq:Hessian_matrix_2}
\end{align}
In \eqref{eq:Hessian_matrix_2}, the terms in first line $2\nabla_{{\bf f}^{\sf H}}\mathcal{L}\left({\bf f}^{\star}\right)\cdot
    \left(
    \sum_{i=1}^{K}\frac{{\bf A}_i{\bf f}^{\star}}{\left(({\bf f}^{\star})^{\sf H}{\bf A}_i{\bf f}^{\star}\right)}-\sum_{i=1}^{K}\frac{{\bf B}_i{\bf f}^{\star}}{\left(({\bf f}^{\star})^{\sf H}{\bf B}_i{\bf f}^{\star}\right)}\right)^{\sf H}$ become zero from the result of Theorem 1. As a result, the extended Hessian matrix reduces to $\nabla_{{\bf f}^{\sf H}}^2\mathcal{L}({\bf f}^{\star})
        = 4\mathcal{L}({\bf f}^{\star})\left(
     \sum_{i=1}^{K}\frac{{\bf B}_i{\bf f}^{\star}\left({\bf f}^{\star}\right)^{\sf H}{\bf B}_i}{(\left({\bf f}^{\star}\right)^{\sf H}{\bf B}_i{\bf f}^{\star})^2}-\sum_{i=1}^{K}\frac{{\bf A}_i{\bf f}^{\star}\left({\bf f}^{\star}\right)^{\sf H}{\bf A}_i}{(\left({\bf f}^{\star}\right)^{\sf H}{\bf A}_i{\bf f}^{\star})^2}
     \right).$ The first term $\mathcal{L}({\bf f}^{\star})$ is a positive scalar and all the remaining terms are the sum of positive-definite matrices. This is due to the fact that ${\bf A}_i$ and ${\bf B}i$ are Hermitian matrices for all $i\in[K]$. If the minimum eigenvalue of $\sum_{i=1}^{K}\frac{{\bf A}_i{\bf f}^{\star}\left({\bf f}^{\star}\right)^{\sf H}{\bf A}_i}{(\left({\bf f}^{\star}\right)^{\sf H}{\bf A}i{\bf f}^{\star})^2}$ is greater than the maximum eigenvalue of $\sum_{i=1}^{K}\frac{{\bf B}_i{\bf f}^{\star}\left({\bf f}^{\star}\right)^{\sf H}{\bf B}_i}{(\left({\bf f}^{\star}\right)^{\sf H}{\bf B}i{\bf f}^{\star})^2}$, then the Hessian matrix $\nabla{{\bf f}^{\sf H}}^2\mathcal.L({\bf f}^{\star},\lambda)$ is guaranteed to be negative-definite. This proves the statement.



\bibliographystyle{IEEEtran}
\bibliography{IEEEabrv,mainOneColumn_Ver10}

\end{document}